\documentclass[aip,jcp,floatfix]{revtex4-1}

\usepackage{graphicx}

\begin{document}

\renewcommand{\vec}[1]{\mathbf{#1}}
\newcommand{\fHxc}{f_{Hxc}}
\newcommand{\fxc}{f_{xc}}
\newcommand{\fxcALDA}{f^\mathrm{ALDA}_{xc}}
\newcommand{\fxcrALDA}{f^\mathrm{rALDA}_{xc}}
\newcommand{\fxcrALDAl}{f^{\mathrm{rALDA},\lambda}_{xc}}
\newcommand{\fxcrALDAt}{f^\mathrm{rALDA}_{xc}}
\newcommand{\fxcrALDAc}{f^\mathrm{rALDAc}_{xc}}
\newcommand{\fxcCDOP}{f^\mathrm{CDOP}_{xc}}
\newcommand{\fxcCDOPt}{f^\mathrm{CDOP}_{xc}}
\newcommand{\fxcCDOPs}{f^\mathrm{CDOPs}_{xc}}
\newcommand{\fxcCP}{f^\mathrm{CP}_{xc}}
\newcommand{\fxcCPd}{f^\mathrm{CPd}_{xc}}
\newcommand{\fxcJGMs}{f^\mathrm{JGMs}_{xc}}
\newcommand{\fxcJGMsl}{f^{\mathrm{JGMs},\lambda}_{xc}}
\newcommand{\fxcJGMst}{f_{xc}^{\mathrm{JGMs}}}
\newcommand{\kF}{k_F}
\newcommand{\veff}{v_\mathrm{eff}}
\def\l{\lambda}
\def\a{\alpha}
\def\w{\omega}
\def\<{\langle}
\def\>{\rangle}

\title{
Adiabatic-connection fluctuation-dissipation DFT
for the structural properties of solids---
the renormalized ALDA and electron gas kernels}

\author{Christopher E. Patrick}
\email[]{chripa@fysik.dtu.dk}
\author{Kristian S. Thygesen}
\email[]{thygesen@fysik.dtu.dk}
\affiliation{Center for Atomic-Scale Materials Design (CAMD), Department of Physics,
Technical University of Denmark, DK---2800 Kongens Lyngby, Denmark}

\date{\today}

\begin{abstract}
We present calculations of the correlation energies of crystalline solids
and isolated systems
within the adiabatic-connection
fluctuation-dissipation formulation of density-functional theory.
We perform a quantitative comparison of a set of model exchange-correlation kernels originally
derived for the homogeneous electron gas (HEG),
including the recently-introduced renormalized
adiabatic local-density approximation (rALDA) and also
kernels which (a) satisfy known exact limits of the HEG,
(b) carry a frequency dependence or (c) display a 1/$k^2$ divergence 
for small wavevectors.
After generalizing the kernels to inhomogeneous systems through a reciprocal-space 
averaging procedure, we calculate the lattice constants and bulk moduli
of a test set of 10 solids consisting of tetrahedrally-bonded semiconductors
(C, Si, SiC), ionic compounds (MgO, LiCl, LiF) and metals (Al, Na, Cu, Pd).
We also consider the atomization energy of the H$_2$ molecule.
We compare the results calculated with different kernels to those obtained from 
the random-phase approximation (RPA) and to experimental
measurements.
We demonstrate that the model kernels correct the RPA's
tendency to overestimate the magnitude of the correlation energy whilst
maintaining a high-accuracy description of structural properties.
\end{abstract}


\maketitle

\section{Introduction}
The remarkable rise of density-functional theory\cite{Hohenberg1964} (DFT) in the last
few decades owes much to the efficient treatment of exchange and correlation
within the local-density approximation (LDA).\cite{Kohn1965}
However, known deficiencies in the LDA's description of
certain systems has led to the development of a hierarchy of exchange-correlation functionals
of varying computational expense.\cite{Becke2014}
At the high-complexity end of this spectrum of functionals lies the adiabatic connection-fluctuation
dissipation formulation of DFT (ACFD-DFT),\cite{Langreth1977,Gunnarsson1976} which in its 
simplest form corresponds to the random-phase approximation (RPA) correlation energy.\cite{Ren2012}
``Beyond-RPA'' methods strive for an even higher level of accuracy and form an important 
and fast-developing field of research.\cite{Ren2012,Marini2006,Ruzsinszky2011,Hesselmann2011,Freeman1977,Paier2010,Niquet2003}

ACFD-DFT provides a natural path for the improvement of the RPA via the introduction
of an exchange-correlation (XC) kernel $\fxc$, an ubiquitous quantity
in time-dependent DFT (TD-DFT).\cite{Runge1984,Petersilka1996}
The homogeneous electron gas (HEG) has become a key system for
the development and testing of new kernels through the ACFD-DFT
calculation of correlation energies.\cite{Lein2000,Constantin2007,
Corradini1998, Richardson1994}
Jellium slabs also form important test systems, 
through the ACFD-DFT calculation of their surface energies,\cite{Pitarke2001,Pitarke2003}
inter-slab interaction energies,\cite{Dobson2000,Jung2004,Constantin2011}
and as benchmarks for widely-used semilocal XC functionals.\cite{Constantin20112,Pitarke2006}
However the last few years have seen the application of the 
ACFD-DFT formalism to calculate the correlation
energies of atoms, molecules and solids, with promising results.\cite{Hellgren2008,Furche2005, Gould20122,Lu2014}

One such example is the ``renormalized kernel approach'', which was introduced based on a model
XC-kernel named the renormalized adiabatic LDA (rALDA).\cite{Olsen2012,Olsen2013}
The rALDA exploits the accurate reciprocal-space description of HEG correlation
provided by the adiabatic LDA (ALDA) in the long-wavelength limit,
whilst correcting the ALDA's unphysical behavior at short wavelengths.
The rALDA and its generalized-gradient analogue (rAPBE) have been shown
to yield highly accurate atomization and cohesive energies
of molecules and solids.\cite{Olsen2012,Olsen2013,Olsen2014}

It is interesting to place the rALDA into the context
of other HEG kernels.
Many theoretical studies have explored the properties
of the exact XC-kernel and derived certain limits which are not necessarily
obeyed by the rALDA.\cite{Ichimaru1982, Holas1987, Gross1985}
Similarly, the rALDA is static, and apart from studies of the HEG\cite{Lein2000}
there is little known about dynamical effects on ACFD-DFT
correlation energies.
Furthermore the XC-kernel of an insulator is known to behave qualitatively 
differently to a metallic system like the HEG,\cite{Aulbur1996}
with the XC-kernel of an insulator famously
diverging $\propto 1/k^2$ in the limit of small wavevectors $k$.\cite{Ghosez1997} 
In this respect it is important to test the validity of 
applying a model HEG kernel to non-metallic systems.

This work explores the above aspects through a quantitative comparison of model
HEG XC-kernels.
Within our sample of XC-kernels we include the rALDA,\cite{Olsen2012} and also 
a kernel which satisfies
exact limits of the HEG,\cite{Corradini1998} a simple
dynamical kernel,\cite{Constantin2007} and a kernel which has
a divergence $\propto 1/k^2$ when describing an insulator.\cite{Trevisanutto2013}
For each XC-kernel we use ACFD-DFT to calculate the correlation energy  of 
a test set of 10 crystalline solids 
and evaluate the lattice constant and bulk modulus,
which we then compare to calculations using semilocal functionals and the RPA, and also
to experiment.
We also provide a demonstrative calculation of the 
atomization energy of the hydrogen molecule to highlight the importance
of spin-polarization.
We find that all of the model XC-kernels greatly improve the magnitude
of the RPA correlation energy whilst providing a highly accurate
description of structural properties.

Our study is organized as follows.
In section~\ref{sec.theory} we review ACFD-DFT and the role played
by the XC-kernel.
In particular, we describe the expected behavior of the XC-kernel for the
HEG at certain limits (section~\ref{sec.kernel_HEG}), introduce
our chosen set of model kernels (section~\ref{sec.kernelzoo})
and apply them to the HEG (section~\ref{sec.corr_hole}).
For inhomogeneous systems we
require a scheme to generalize HEG kernels for a varying density;
in section~\ref{sec.averaging} we discuss possible schemes and 
justify the choice made in this work.
Section~\ref{sec.results} contains the results of our study,
in which we discuss the calculated lattice constants and bulk
moduli, absolute correlation energies and 
the H$_2$ molecule.
Finally in section~\ref{sec.conclusions} we summarize our results
and offer our conclusions.

\section{Theory}
\label{sec.theory}

\subsection{Correlation energies in the ACFD-DFT framework}

Here we summarize the essential concepts of ACFD-DFT.
Full derivations may be found in original articles\cite{Langreth1977,Gunnarsson1976}
or recent reviews, e.g.\ Refs.~\citenum{Ren2012}~and~\citenum{Leinbook}. 

In ACFD-DFT, a system of fully-interacting
electrons is described by a coupling-constant dependent Hamiltonian $H(\l)$.
The coupling constant $\l$ takes values between 0 and 1 and 
defines an effective interaction between 
electrons as $\l v_c$,
where $v_c$ is the Coulomb interaction.
$H(\l=1)$ corresponds to the exact Hamiltonian of the fully-interacting system.
In addition to the effective Coulomb interaction $H(\l)$ 
contains a $\l$-dependent single-particle potential $v^\l_\mathrm{ext}$,
constructed in such a way that the ground-state 
solution of $H(\l)$ has exactly
the same electronic density as the ground-state solution of the 
fully-interacting ($\l = 1$) Hamiltonian.
The fixed-density path connecting $\l=0$ and 1 defines the ``adiabatic connection''.
Since $H(\l = 0)$ describes a system of non-interacting electrons
with a fully-interacting density,
$v^{\l=0}_\mathrm{ext}$ is readily identified as the Kohn-Sham 
potential from DFT.\cite{Hohenberg1964, Kohn1965}

Invoking the Hellmann-Feynman theorem, integrating with respect to $\l$ along 
the adiabatic connection and comparing to standard DFT\cite{Hohenberg1964} 
yields an expression for the 
exchange-correlation (XC) energy in terms of the operator
describing density fluctuations.
The fluctuation-dissipation theorem\cite{Pinesbook}
provides the link between this operator and
the frequency integral
of a response function $\chi^\l$.
For non-interacting electrons, $\chi^{\l=0} \equiv \chi_\mathrm{KS}$, the Kohn-Sham response
function of time-dependent DFT.\cite{Runge1984,Petersilka1996}
The XC-energy is then written  as the sum of an ``exact'' exchange contribution $E_\mathrm{x}$ and 
the correlation energy $E_\mathrm{c}$,  where the latter is given by (in Hartree
units):
\begin{equation}
E_\mathrm{c} = -\frac{1}{2\pi} \sum_\vec{q} \int_0^1 d\l \int_0^\infty ds  \ 
\mathrm{Tr}\left[v_c(\vec{q})(\chi^\l(\vec{q},is) - 
\chi_\mathrm{KS}(\vec{q},is)) 
\right].
\label{eq.Ec}
\end{equation}
Equation~\ref{eq.Ec} has been written in a plane-wave basis so that 
the quantities on the right-hand side are matrices in the reciprocal 
lattice vectors $\vec{G}$ and $\vec{G'}$, and the wavevectors $\vec{q}$
belong to the first Brillouin zone.
The Coulomb interaction is diagonal in a plane-wave 
representation with elements $4\pi/|\vec{q}+\vec{G}|^2$,
and $s$ is a real number corresponding to an imaginary frequency, $\w = is$.

The link between the interacting and non-interacting response functions
is supplied by linear-response theory,\cite{Petersilka1996} which
describes the behavior of density $n$ in the presence of a small perturbation:
\begin{equation}
\delta n (\vec{q},\w) = \chi^\l(\vec{q},\w) \delta v^\l_\mathrm{ext}(\vec{q},\w).
\label{eq.dens_pert}
\end{equation}
The fact that $\chi^\l$ yields the exact density response
at all values of $\l$ allows the link to be made to $\chi_\mathrm{KS}$
through the following integral equation,
\begin{equation}
\chi^\l(\vec{q},\w) = \chi_\mathrm{KS}(\vec{q},\w) + 
 \chi_\mathrm{KS}(\vec{q},\w)
\fHxc^\l(\vec{q},\w)
\chi^\l(\vec{q},\w)
\label{eq.chi_ks}
\end{equation}
where the Hartree-XC kernel $\fHxc^\l$ has been introduced as
\begin{equation}
\fHxc^\l(\vec{q},\w) = \l v_c(\vec{q}) + \fxc^\l(\vec{q},\w).
\end{equation}
The XC-kernel $\fxc^\l$ describes the change in the XC-potential $v^\l_{xc}$
upon perturbing the density, which is a fully nonlocal quantity
in time and space:\cite{Petersilka1996}
\begin{equation}
\fxc^\l(\vec{r},\vec{r'},t-t') = \frac{\delta v^\l_{xc}(\vec{r},t)}{\delta n(\vec{r'},t')}
\label{eq.fxc}
\end{equation}
By assuming approximate forms for $\chi_\mathrm{KS}$ and 
$\fxc^\l$, one can use equation~\ref{eq.chi_ks} to calculate
$\chi^\l$ 
and thus evaluate the correlation energy with equation~\ref{eq.Ec}.

\subsection{ACFD-DFT in practice}

In a plane-wave basis, the Kohn-Sham response function 
has the form\cite{Yan2011}
\begin{equation}
\chi_\mathrm{KS}^{\vec{G}\vec{G'}}(\vec{q},is) = \frac{2}{\Omega}
\sum_{\vec{k}\nu\nu'}(f_{\nu\vec{k}} - f_{\nu'\vec{k}+\vec{q}}) \
\frac{
n_{\nu\vec{k},\nu'\vec{k}+\vec{q}}(\vec{G})
n^*_{\nu\vec{k},\nu'\vec{k}+\vec{q}}(\vec{G'})
}
{is + \varepsilon_{\nu\vec{k}} - \varepsilon_{\nu\vec{k}+\vec{q}}},
\label{eq.chiPW}
\end{equation}
where $f_{\nu\vec{k}}$ and $\varepsilon_{\nu\vec{k}}$ represent the occupation
factor
and energy of the Kohn-Sham state $\psi_{\nu\vec{k}}$, while
the pair densities
$n_{\nu\vec{k},\nu'\vec{k}+\vec{q}}(\vec{G})$ are matrix elements of plane waves,
$\<\psi_{\nu\vec{k}}| e^{-i(\vec{q}+\vec{G})\cdot\vec{r}} | \psi_{\nu'\vec{k}+\vec{q}}\>$.
$\Omega$ is the volume of the primitive unit cell, and the factor
of 2 assumes a spin-degenerate system.
From equations~\ref{eq.Ec}~and~\ref{eq.chiPW} the benefits of the ACFD-DFT are
not very obvious; to construct
$\chi_\mathrm{KS}$ we require $\psi$, which means solving
the Kohn-Sham equations and thus already obtaining the correlation energy.
Furthermore, to solve the integral equation (\ref{eq.chi_ks}) we
require the XC-kernel $\fxc$, 
which arguably is even more complicated  than the XC-potential $v_{xc}$
due to its frequency dependence.

However the attraction of ACFD-DFT is that
even setting $\fxc=0$  yields both 
a nonlocal description of exchange and a nontrivial expression
for the correlation energy, namely that obtained from the RPA:\cite{Ren2012}
\begin{equation}
E^\mathrm{RPA}_\mathrm{c} = \frac{1}{2\pi} 
\sum_\vec{q} \int_0^\infty ds \ \mathrm{Tr}\left[
\mathrm{ln}\left\{1 - v_c(\vec{q})\chi_\mathrm{KS}(\vec{q},is)\right\} \right . \
 \left. +
v_c(\vec{q})\chi_\mathrm{KS}(\vec{q},is)
\right]   .
\label{eq.EcRPA}
\end{equation}
The RPA has been applied across a wide range of physical 
systems\cite{Furche2001,Fuchs2002,Lu2009,Nguyen2009,Harl2010,Schimka2010, Olsen2011, Olsen20132}
and found to give a markedly improved description
of nonlocal correlation effects.
Equation~\ref{eq.EcRPA} is usually applied 
as a post-processing step to a
DFT calculation, analogous to $G_0W_0$ corrections to band gaps.\cite{Hedin1970,Hybertsen1986}

Based on the success of the RPA it may be hoped
that the description of correlation might be further improved by using more sophisticated approximations for $\fxc$.
While it turns out that the adiabatic local-density approximation (ALDA)
offers little improvement,\cite{Furche2005} nonlocal, dynamical and/or
energy-optimized approximations for $\fxc$  have been found to 
correct deficiencies of the RPA when calculating the
correlation energy of the homogeneous electron gas 
(HEG).\cite{Richardson1994, Corradini1998, Dobson2000, Lein2000, Jung2004, Constantin2007, Pitarke2001,
Pitarke2003, Constantin2011}

We note that a non-self-consistent
application of the ACFD-DFT formula might suffer from a dependence
on $v_{xc}$, the exchange-correlation potential used to construct
the orbitals forming $\chi_\mathrm{KS}$.
In this respect, a self-consistent scheme is
attractive and the subject of current 
research.\cite{Godby1988, Kotani1998, Nguyen2014,Klimes2014}
However here we do not include any self-consistency and
treat $\fxc$ as a quantity to be optimized independent of the 
$v_{xc}$ used to generate $\chi_\mathrm{KS}$.\cite{consnote}

\subsection{XC-kernels from the homogeneous electron gas}
\label{sec.kernel_HEG}

In the same way that the HEG is used to generate
approximate XC-potentials, it also forms a natural
starting point for approximate XC-kernels.
Here we review some properties of the exact XC-kernel
of the HEG.

\subsubsection{Definitions}

The analogue of equation~\ref{eq.chi_ks} for the
HEG is:
\begin{equation}
\chi^\l(k,\w) = \chi_0(k,\w) + 
\chi_0(k,\w)
\fHxc^{\mathrm{HEG},\l}(k,\w)
\chi^\l(k,\w).
\label{eq.chi_HEG}
\end{equation}
All quantities appearing in this equation are scalars, with $ k = |\vec{G} + \vec{q}|$. 
$\chi_0$ is the Lindhard response function, with occupation numbers equal to 1 for
plane-wave states below the Fermi level and zero otherwise.\cite{Inksonbook}
As demonstrated in the appendix of Ref.~\citenum{Farid1993}, in the case
that equation~\ref{eq.chi_ks} is applied to the HEG 
the Lindhard and Kohn-Sham
response functions coincide, so that
the quantity $\fHxc^{\mathrm{HEG},\l}$  
appearing in equation~\ref{eq.chi_HEG} must also match
its counterpart $\fHxc^\l$ in equation~\ref{eq.chi_ks}.
Therefore
the HEG forms a rigorous test ground for model XC kernels.

For simplicity we denote $\fxc^{\mathrm{HEG},\l=1}$ as
$\fxc^\mathrm{HEG}$.
The local field factor $G(k,\w)$ is
related to $\fxc^{\mathrm{HEG}}(k,\w)$ as $\fxc^{\mathrm{HEG}}(k,\w)= -v_c(k) \ G(k,\w)$,
so that
equation~\ref{eq.chi_HEG} defines $G$ in terms of
the response functions  $\chi^{\l = 1}$ and $\chi_0$.\cite{Holas1987}
A potential source of confusion is that
another field factor $G_I$ can be found in the 
literature which has a different defining equation.\cite{Niklasson1974,Vignale1988,Richardson1994,Holas1987}
For $G_I$, 
the Lindhard response function appearing in equation~\ref{eq.chi_HEG} is 
modified such that the occupation numbers of each plane-wave state are
calculated using the many-body wavefunction of the fully-interacting 
system.\cite{Niklasson1974}
Therefore $\fxc^{\mathrm{HEG}, I}$ is also a distinct quantity.
However the kernels investigated in this work were derived based on the 
equivalence of 
equations~\ref{eq.chi_ks}~and~\ref{eq.chi_HEG} for the HEG,\cite{Farid1993}
so it is $G$ which  is of interest in the current work.

\subsubsection{Exact limits}
\label{sec.limits}

The local-field factor $G$ (and thus $\fxc^\mathrm{HEG}$)
have been the subject
of many theoretical studies (c.f.\ section IIIC of Ref.~\citenum{Ichimaru1982}),
and their behavior  at certain limits is known exactly.
First, in the long wavelength ($k \rightarrow 0$) 
and static ($\w =0 $) limit, the HEG XC-kernel reduces
to the adiabatic local-density approximation (ALDA):
\begin{equation}
\fxc^\mathrm{HEG}(k\rightarrow 0, \w = 0) = \fxcALDA
\equiv -\frac{4 \pi A}{k_F^2} 
\label{eq.smallq}
\end{equation}
where
\begin{equation}
A = \frac{1}{4} - \frac{k_F^2}{4\pi}\frac{d^2 (n\varepsilon_c)}{dn^2}.
\label{eq.A}
\end{equation}
$k_F = (3\pi^2n)^{1/3}$ is the Fermi wavevector for the HEG of density $n$, and $\varepsilon_c$
is the correlation energy per electron.
The two terms in equation~\ref{eq.A} correspond to the exchange and
correlation contributions to the ALDA kernel.
Equation~\ref{eq.smallq} can be seen either as a consequence of the compressibility
sum rule,\cite{Ichimaru1982} or more simply by noting that 
the ALDA should be exact in describing the HEG response to a uniform,
static field.\cite{Constantin2007}

Remaining in the static case, but considering small wavelengths
($k \rightarrow \infty$) yields\cite{Holas1987,Toulouse2005}
\begin{equation}
\fxc^\mathrm{HEG}(k\rightarrow\infty, \w = 0) = -\frac{4\pi B}{k^2} - \frac{4\pi C}{k_F^2} 
\label{eq.bigq}
\end{equation}
whilst in the long wavelength, high frequency limit 
($k = 0, \w \rightarrow \infty $) $\fxc^\mathrm{HEG}$ is given by\cite{Gross1985}
\begin{equation}
\fxc^\mathrm{HEG}(k=0, \w \rightarrow \infty) = -\frac{4 \pi D}{k_F^2}.
\label{eq.highfreq}
\end{equation}

Although we have not written it explicitly, $A$, $B$, $C$ and $D$ depend
on the density of the HEG, or equivalently on the Fermi wavevector or Wigner radius
$r_s = (3/4\pi n )^{1/3}$.
Practically $A$, $C$ and $D$ can be obtained from a parameterization of the
HEG correlation energy $\varepsilon_c$, while $B$ 
additionally requires the momentum distribution and on-top
pair-distribution function of the HEG.\cite{Ortiz1994}
In this work we use the parameterization of $\varepsilon_c$ and $B$
from Refs.~\citenum{Perdew1992}~and~\citenum{Moroni1995} respectively.

\subsubsection{Intermediate $k$ values}

In addition to these limits, calculating the correlation energy
from equation~\ref{eq.Ec} requires an expression for $\fxc^\mathrm{HEG}$ across
all $k$ and $\w$.
Ref.~\citenum{Moroni1995} provided important insight into the HEG XC-kernel
with diffusion Monte Carlo calculations of $\fxc^\mathrm{HEG}(k,\w=0)$ for a range
of $k$ vectors and densities.
Interestingly, one of the conclusions of the work was that in this
static limit and for $k\lesssim 2k_F$,  $\fxc^\mathrm{HEG}$ 
could be well-approximated by taking its $k=0$ limiting
form (i.e.\ the ALDA, equation~\ref{eq.smallq}).

\subsubsection{The large-k limit and the pair-distribution function}
\label{sec.paircorrelation}

According to equation~\ref{eq.bigq}, the $k\rightarrow\infty, \w=0$ limit of
$\fxc^\mathrm{HEG}$ is a constant, and thus
$G(k,\w=0)$ diverges as $k^2$.
This property appears to have alarming consequences for the pair-distribution function: 
in  Ref.~\citenum{Kimball1973} it is shown that
a $G(k)$ which diverges for large $k$ will yield a singular pair-distribution function
at the origin (see also Refs.~\citenum{Singwi1970,Furche2005}).
The crucial point to note however is that the relationship in Ref.~\citenum{Kimball1973}
assumes a \emph{frequency-independent} $G(k)$, which is not the same as the \emph{frequency-dependent
$G(k,\w)$ evaluated at $\w = 0$}, which equation~\ref{eq.bigq} describes.
This point is discussed further in Ref.~\citenum{Farid1993}.

Nonetheless, a frequency-independent model kernel satisfying
the exact HEG limits for small and large $k$ at $\w = 0$ (e.g.\ the CDOP
kernel\cite{Corradini1998}) will have a badly-behaved
pair-distribution function, as demonstrated below and
in Fig.~2 of Ref.~\citenum{Lu2014}.
In this respect, and given that the correlation energy is calculated
as an integral over frequency, a frequency-averaged 
$G(k)$ may be a better starting point for model kernels than $G(k,\w=0)$.
Below we compare kernels which do or do not satisfy equation~\ref{eq.bigq}.

\subsection{Model XC kernels}
\label{sec.kernelzoo}

\begin{figure*}
\includegraphics[width=120mm]{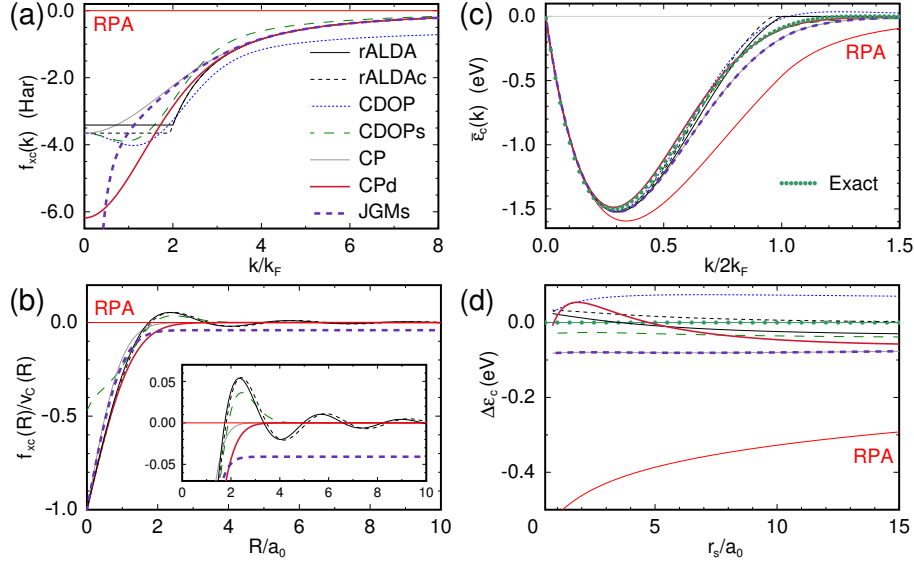}
\caption{
Comparison of model HEG XC-kernels, plotted in (a) reciprocal
and (b) real space for $r_s=2$.
In (b) the XC kernels are divided by the Coulomb
interaction $v_c$, and the inset provides a zoomed image close
to $\fxc = 0$.
Apart from the JGMs all the XC-kernels are short-range,
while (apart from CDOPs) at small R the XC-kernels cancel the Coulomb
interaction such that $\fHxc$ vanishes.
We have omitted the CDOP kernel in (b), since it matches CDOPs apart from
a $\delta$-function at $R=0$ (equation~\ref{eq.CDOPreal}).
The CPd kernel was evaluated at an energy of 2 Hartrees
and the JGMs kernel at a band gap of 3.4~eV.
In (c) we plot the wavevector-resolved correlation energy
(equation~\ref{eq.k_resolved}) at $r_s=4$ compared to the 
exact\cite{Lein2000} result
obtained from the parameterization of the correlation
hole given in Refs.~\citenum{Perdew19922}~and~\citenum{Perdew1997}.
In (d) we plot the difference in calculated correlation energies
with a parameterization\cite{Perdew1992}
of Monte Carlo calculations\cite{Ceperley1980} of $\varepsilon_c$.
\label{fig.kernels}
}
\end{figure*}

Having introduced the relevant quantities for the HEG,  we
now describe the XC-kernels investigated in the current work.
The XC-kernels are plotted in reciprocal or real
space  in Figs.~\ref{fig.kernels}(a)~and~(b).
The reciprocal and real space XC-kernels
are related by the Fourier transform
\begin{equation}
\fxc(n,k,\w) = \int d\vec{R} e^{-i\vec{k}\cdot\vec{R}} \fxc(n,R,\w)
\end{equation}
with $R = |\vec{r} - \vec{r'}|$.

\subsubsection{The rALDA kernel}

The renormalized adiabatic local density approximation (rALDA)\cite{Olsen2012}
XC-kernel is given by
\begin{equation}
\fxcrALDA(n,k) =  - \left[ \theta(k_c-k) \frac{4\pi}{k_c^2} + \theta(k-k_c) \frac{4\pi}{k^2} \right]
\label{eq.ralda}
\end{equation}
where the Heaviside function $\theta(x) = 1$ for $x>0$ and zero otherwise.
The cutoff wavevector is chosen as
\begin{equation}
k_c = \kF/\sqrt{A}
\label{eq.ralda_A}
\end{equation} 

In previous applications of this kernel\cite{Olsen2012,Olsen2013} 
the coefficient $A$ defined by equation~\ref{eq.A}
was replaced by  $1/4$, corresponding to omitting the correlation contribution.
In this work we shall refer to this exchange-only kernel
as the rALDA kernel $\fxcrALDA$.
The special label $\fxcrALDAc$ (rALDAc) refers to the kernel calculated including 
both the exchange and correlation contributions in equation~\ref{eq.A}.
We note that the rALDAc kernel coincides with that of Ref.~\citenum{Toulouse2005}
with $B=1$ and $C=0$.

$\fxcrALDAc$ obeys the exact $k\rightarrow 0 $, $\w=0$ limit for the HEG
(equation \ref{eq.smallq}).
Furthermore, both $\fxcrALDAc$
and $\fxcrALDA$ mimic the HEG kernel\cite{Moroni1995} in 
displaying small variation for wavevectors below $\sim 2\kF$.
At larger wavevectors both kernels correspond exactly to the Coulomb interaction with opposite sign,
such that the corresponding Hartree-XC kernels $\fHxc$ vanish for $k>k_c$.
In real space the kernel has the form
\begin{equation}
\fxcrALDAt(n,R) = -\frac{1}{R} \left[ 1 - \frac{2}{\pi}\left(\int_0^{k_cR} \frac{\sin x}{x} dx  \right. \right. 
\left.\left. -  \frac{ [\sin(k_cR)  -  k_c R \cos(k_c R)]}{(k_c R)^2}\right)\right]
\end{equation}
with the Fourier transform of the Heaviside functions leading to decaying
oscillations [Fig.~\ref{fig.kernels}(b)].  
At small $R$ the XC-kernel diverges as $-1/R$, yielding
a Hartree-XC kernel which is finite at the origin.\cite{Olsen2012}

\subsubsection{The CDOP kernel}
The kernel introduced by Corradini, del Sole, Onida and Palummo (CDOP)
in Ref.~\citenum{Corradini1998} has the form
\begin{equation}
\fxcCDOP(n,k) =
-\left[\frac{4\pi \alpha}{\kF^2} \left(\frac{k}{\kF}\right)^2 e^{-\beta(k/\kF)^2}  \right. 
\left. 
+ \frac{4\pi B}{\kF^2} \frac{1}{[g + (k/\kF)^2]}
+\frac{4\pi C}{\kF^2} \right]
\label{eq.cdop}
\end{equation}
where $g = B/(A-C)$, and $\alpha$ and $\beta$ are density-dependent 
fitting parameters chosen to best reproduce the local field factor
$G(k,\w=0)$ obtained from the QMC calculations of Ref.~\citenum{Moroni1995}.

Uniquely among the kernels considered here, the CDOP kernel obeys 
both the $k\rightarrow 0$ and $k\rightarrow\infty$ limits of the HEG
at $\w = 0$, equations \ref{eq.smallq} and \ref{eq.bigq}.
However as noted above the short-wavelength $C$ term causes the pair-distribution function
to diverge.\cite{Lu2014}
In Ref.~\citenum{Lu2014} a simplified kernel which avoids this divergence
was obtained from equation~\ref{eq.cdop} by setting $C=0$.
We shall also investigate this kernel in this work, labeled CDOPs ($\fxcCDOPs$).

The real-space form of the CDOP kernel is:
\begin{equation}
\fxcCDOPt (n, R)=  -\frac{1}{R} B e^{-\sqrt{g}k_F R} - \frac{4\pi C}{k_F^2} \delta(R)
+ \frac{\a k_F}{4\pi^2\beta} \left(\frac{\pi}{\beta}\right)^\frac{3}{2} 
\left[ \frac{k_F^2 R^2}{2\beta} - 3\right] e^{-k_F^2R^2/4\beta} 
\label{eq.CDOPreal}
\end{equation}
We note that the $C$ term in equation~\ref{eq.cdop} produces a $\delta$-function
in real-space, while for small $R$ (excluding the $\delta$-function) the XC-kernel
diverges as $-B/R$, such that the Hartree-XC kernel is still divergent
as $(1-B)/R$.

\subsubsection{The CP kernel}

A kernel with a simpler functional form was introduced by Constantin and Pitarke (CP)
in Ref.~\citenum{Constantin2007}:
\begin{equation}
\fxcCP(n,k) = -\left[\frac{4\pi}{k^2} (1 - e^{-\kappa_0 k^2})\right]
\label{eq.CP}
\end{equation}
Here $\kappa_0 = A/\kF^2$, which ensures that the HEG
$k\rightarrow0$, $\w = 0$ limit is satisfied.
Like the rALDA kernels, at large wavevectors $\fxcCP(k)$ 
cancels the Coulomb interaction so that 
$\fHxc$ vanishes.
The CP kernel possesses a compact form in real space 
in terms of the error function:
\begin{equation}
\fxcCP(n,R) = - \frac{1}{R} \left[ 1 - 
\mathrm{erf} \left(\frac{R}{\sqrt{4\kappa_0}}\right)\right].
\label{eq.CPreal}
\end{equation}
As $R\rightarrow 0$, $\fxcCP$ diverges as $-1/R$ and thus yields a finite
Hartree-XC kernel
in this limit.

\subsubsection{The CPd dynamical kernel}

With its simple form, the CP kernel is an ideal starting point to explore
more complex aspects of $\fxc$, such as its frequency dependence.
In Ref.~\citenum{Constantin2007}, a  dynamical kernel was introduced
(CPd) by replacing $\kappa_0$ appearing in equation~\ref{eq.CP}
with $\kappa_\w$, i.e.
\begin{equation}
\fxcCPd(n, k, \w) = -\left[\frac{4\pi}{k^2} (1 - e^{-\kappa_\w k^2})\right]
\label{eq.CPd}
\end{equation}
where for imaginary frequency $\w = is$, 
\begin{equation}
\kappa_\w = \kappa_0 \frac{1 + a s + c s^2}{1 + s^2}.
\end{equation}
In Ref.~\citenum{Constantin2007} the coefficient $c = D/A$ was chosen to 
correctly reproduce the $k\rightarrow 0$, $\w \rightarrow \infty$ 
limit of the HEG (equation~\ref{eq.highfreq}), while 
the relation $a = 6\sqrt{c}$ was found to give a good fit to the 
correlation energy calculated for the HEG using $\fxcCPd$.
We note that the CPd kernel varies non-monotonically with frequency
in the $k\rightarrow 0$ limit.\cite{Constantin2007}

\subsubsection{The JGMs kernel}

The limits of the exact kernel discussed in section~\ref{sec.limits}
were derived for the HEG, which is metallic.
However, the XC-kernel of a periodic \emph{insulator}
is known to behave differently,
diverging $\propto$ $1/k^2$ in the $k\rightarrow0$
limit.\cite{Aulbur1996,Ghosez1997}
This limit has been found to play an essential role in the TD-DFT calculation of 
excitonic effects in optical spectra, leading to the development
of kernels which exhibit the same $1/k^2$  
divergence.\cite{Botti2004, Sharma2011,Trevisanutto2013}
Here we focus on the ``jellium with gap'' model (JGM) kernel of 
Ref.~\citenum{Trevisanutto2013},
which has the useful property of reducing to a model HEG 
kernel\cite{Constantin2007}
when applied to a metallic system.

The JGM kernel of Ref.~\citenum{Trevisanutto2013} was derived
based on two steps.
First, the theoretical arguments of Ref.~\citenum{Sottile2003}
were used to connect the $k\rightarrow0$ limit of $\fxc$ to
$\epsilon$, the dielectric function, 
as $\fxc(k\rightarrow 0) = -4\pi/[k^2(\epsilon - 1)]$
(a similar relation was found empirically in Ref.~\citenum{Botti2004}).
Then, the model dielectric function of Ref.~\citenum{Levine1982}
was used to relate $\epsilon$ to the band gap of the material $E_g$,
as $\epsilon - 1 = 4\pi n/E_g^2$.
The same power dependence may be found for other model dielectric
functions, e.g.\ the Penn model,\cite{Penn1962} and essentially
follows from the $f$-sum rule.
Combining these relations places a requirement on the model kernel 
that it diverges
as $-\alpha /k^2$ in the small $k$ limit, where
$\alpha \rightarrow E_g^2/n$.

In Ref.~\citenum{Trevisanutto2013} the JGM kernel was constructed
to satisfy this divergence, based on a modified CP kernel
introduced in Ref.~\citenum{Constantin2007}.
Following their approach, starting from the unmodified CP kernel
$\fxcCP$ we introduce a simplified JGM (JGMs) kernel $\fxcJGMs$ 
for a system with a gap as
\begin{equation}
\fxcJGMs(n, k, E_g) = -\left[\frac{4\pi}{k^2} (1 - e^{-\kappa_0 k^2} e^{-E_g^2/(4\pi n)} )\right].
\label{eq.JGMs}
\end{equation}
$\fxcJGMs$ has many of the properties of the JGM kernel introduced in 
Ref.~\citenum{Trevisanutto2013}.
For systems with a band gap, the XC-kernel diverges as 
$-\a/k^2$ at small $k$.
For $E_g = 0$, $\fxcJGMs$ reduces to $\fxcCP$
(equation~\ref{eq.CP}), while for $E_g\rightarrow\infty$
$\fxcJGMs \rightarrow -v_c$, yielding a vanishing correlation energy.
Indeed the JGMs kernel differs only from the JGM kernel in its behavior at large $k$,
with the JGM kernel correctly reproducing the $\w=0$ limit of the HEG for $E_g = 0$ 
(equation~\ref{eq.bigq}); concomitantly the JGM kernel
has a diverging pair-distribution function.
By introducing the JGMs kernel we can
study the effect of the $-\a/k^2$ divergence
without any additional complications potentially arising from a
badly-behaved pair-distribution function.

The real-space form of the JGMs kernel is
\begin{equation}
\fxcJGMst(R) = - \frac{1}{R}  \left[1 - e^{-E_g^2/(4\pi n)} 
\ \mathrm{erf}\left(\frac{R}{\sqrt{4\kappa_0}}\right)\right].
\label{eq.JGMR}
\end{equation}
Equation~\ref{eq.JGMR} and Fig.~\ref{fig.kernels}(b) emphasize a unique 
property of the JGMs kernel: it is long range.
As a result,  
at large $R$ the Hartree-XC kernel does not reduce to the bare Coulomb
kernel but rather to an interaction weakened by a factor $\exp[-E_g^2/(4\pi n)]$.

\subsubsection{Coupling-constant dependence}
\label{sec.lambda}

Evaluating the integral over $\l$ in equation~\ref{eq.Ec}
requires the $\fxc$ kernel at an arbitrary coupling strength.
We use the analysis of Ref.~\citenum{Lein2000} to link $\fxc^\l$ to
the fully-interacting kernel
through the relation
\begin{equation}
\fxc^\l(n,k,\w) = \l^{-1} \ \fxc(n/\l^3,k/\l,\w/\l^2).
\label{eq.scaling}
\end{equation}
The scaling of the density can be equivalently
stated as $\l r_s$ or $k_F/\l$.
The (exchange-only) rALDA XC-kernel has the useful property that $\fxcrALDAl = \l\fxcrALDA$.

For the JGMs kernel, we have an additional parameter $E_g$.
For simplicity, we employ a scaling
\begin{equation}
\fxcJGMsl(n,k,E_g) = \l^{-1} \ \fxcJGMs(n/\l^3,k/\l , E_g/\l^{1.5}),
\label{eq.gapscaling}
\end{equation}
equivalent to treating $E_g^2/n$ independent of $\lambda$.

\subsubsection{Analogy with range-separated RPA}
\label{sec.rangeRPA}
It is interesting to draw comparisons
with RPA methods based on the concept of range-separation.\cite{Toulouse2009,Janesko2009,Bruneval2012}
First we 
trivially relabel the Hartree-XC kernel as an effective interaction $\veff$,
i.e.\ $\veff = v_c + \fxc$, 
noting from Fig.~\ref{fig.kernels}(b) that for most of the kernels 
$\veff$ goes to zero at $R \rightarrow 0$ and 
tends to the full Coulomb
interaction at large $R$.
Now, specializing to a static model XC-kernel which
scales linearly with coupling constant, $\fxc^\l(\vec{q}) = \l \fxc(\vec{q})$,
we can partition the correlation energy in equation~\ref{eq.Ec} 
into two contributions
$E_\mathrm{c} = E_\mathrm{c}^\mathrm{LR} + E_\mathrm{c}^\mathrm{SR}$,
with
\begin{equation}
E_\mathrm{c}^\mathrm{LR} = \frac{1}{2\pi} 
\sum_\vec{q} \int_0^\infty ds \ \mathrm{Tr}\left[
\mathrm{ln}\left\{1 - \veff(\vec{q})\chi_\mathrm{KS}(\vec{q},is)\right\} \right.
\left. +
\veff(\vec{q})\chi_\mathrm{KS}(\vec{q},is)
\right]   
\label{eq.Ec_LR}
\end{equation}
\begin{equation}
E_\mathrm{c}^\mathrm{SR} =
\frac{1}{2\pi} \sum_\vec{q} \int_0^1 d\l \int_0^\infty ds  \
\mathrm{Tr}\left[\fxc(\vec{q})(\chi^\l(\vec{q},is) - 
\chi_\mathrm{KS}(\vec{q},is)) 
\right].
\label{eq.Ec_SR}
\end{equation}
Equation~\ref{eq.Ec_LR} matches the RPA expression for
the correlation energy (equation~\ref{eq.EcRPA}) with
$v_c$ replaced with $v_\mathrm{eff}$.
Similarly, equation~\ref{eq.Ec_SR} matches
the expression for the full correlation energy (equation~\ref{eq.Ec})
with $v_c$ replaced by $-\fxc$, which
is generally a short-ranged interaction.

We can focus further on the specific example of the CP XC-kernel (equation~\ref{eq.CP}), 
noting that this kernel can be linearized in $\l$ by neglecting the correlation
contribution to $A$ in equation~\ref{eq.A}.
Then,  $\veff = \mathrm{erf}(\mu R)/R$, with the ``range-separation parameter'' $\mu$ 
determined by the density through $\mu = k_F \sim 1.9/r_s$.
This effective interaction is often found in the range-separated
RPA\cite{Toulouse2009} with $\mu$ of order unity.

We stress that equations~\ref{eq.Ec_LR}~and~\ref{eq.Ec_SR} are 
exact 
for any kernel which obeys 
$\fxc^\l(\vec{q}) = \l \fxc(\vec{q})$.
Since most of the XC-kernels under study here do not obey this relation,
we have not explored equations~\ref{eq.Ec_LR} and \ref{eq.Ec_SR} further
in this work.
However for XC-kernels linear in $\l$ (e.g.\ the rALDA)
there may be a computational advantage in calculating
$E_\mathrm{c}^\mathrm{LR}$ and $E_\mathrm{c}^\mathrm{SR}$
separately.
$E_\mathrm{c}^\mathrm{LR}$  requires only $\chi_\mathrm{KS}$ and not
$\chi^\l$, so for a given kernel it can be obtained 
at exactly the same cost as the RPA correlation energy.
In fact since the effective interaction $\veff$ generally vanishes at large wavevectors,
$E_\mathrm{c}^\mathrm{LR}$ can be expected to avoid the basis-set convergence
problems of the RPA recently highlighted in Ref.~\citenum{Klimes20142}.
Meanwhile it may be possible to exploit the short-range character
of $\fxc$ to reduce the computational cost of calculating $E_\mathrm{c}^\mathrm{SR}$
from equation~\ref{eq.Ec_SR}.

\subsection{Calculating HEG correlation energies}
\label{sec.corr_hole}

A standard test of model HEG kernels is to calculate the 
correlation energy per electron $\varepsilon_c$
from equations~\ref{eq.Ec}~and~\ref{eq.chi_HEG}.
For a given density $r_s$,
$\varepsilon_c$ can be resolved as an integral over $k$ as\cite{Lein2000}
\begin{equation}
\varepsilon_c = \int_0^\infty \bar\varepsilon_c(k) d(k/2k_F).
\label{eq.k_resolved}
\end{equation}
The quantity $\bar\varepsilon_c(k)$ can be compared to the Fourier
transform of a suitable parameterization of the ``exact'' 
correlation hole obtained from Monte Carlo calculations.\cite{Lein2000,Perdew19922,Perdew1997}
Alternatively one can compute $\varepsilon_c$ over a range
of densities and compare to the parameterized result.\cite{Perdew1992}
These comparisons are made in Figs.~\ref{fig.kernels}(c)~and(d)
respectively.

The HEG analysis has been performed a number of times\cite{Lein2000,
Constantin2007,Olsen2012} so we only summarize the key
points.
The RPA correlation is too negative, while including any
of the XC-kernels brings $\varepsilon_c$ to within 0.1~eV of the exact result.
Considering the wavevector decomposition in 
Fig.~\ref{fig.kernels}(c) we find that 
the dynamical CPd kernel provides the best description of 
the correlation hole at this density ($r_s = 4$), but
below $\sim 1.5\kF$
there is very little difference between any of the XC-kernels
and the exact result.
Indeed the ALDA (not shown) also
provides a good description of the correlation hole at these wavevectors.
At larger wavevectors, differences begin to emerge between the kernels,
with the CP kernel becoming too negative, the CPd and CDOPs kernels
closely following the exact result, and the other kernels too positive.
The rALDA kernels are abruptly cut off at $\bar\varepsilon_c(k) = 0$,
while the CDOP kernel acquires a slowly-decaying positive contribution.
The latter behavior is observed to a greater extent in the ALDA,
and originates from the locality of the kernels.\cite{Lein2000,Olsen2012}

Over the full range of densities [Fig.~\ref{fig.kernels}(d)], we find that the calculated correlation energy
is slightly too positive with the CDOP kernel and too negative with the CP and CDOPs
kernels.
Interestingly, CDOPs is closer to the exact result than CDOP,
illustrating that removing the part of the CDOP kernel which causes
the pair-distribution function to diverge\cite{Lu2014} slightly
improves the correlation energy.
The rALDA kernels fall closest to the exact result across a wide range
of densities, and the CPd kernel also provides a good description
of the correlation energy.
Comparing rALDA and rALDAc, we see that removing the
correlation contribution from $A$ in equation~\ref{eq.A} decreases
the correlation energy per electron by less than $\sim$0.02~eV across
a range of densities.

\subsection{Coupling-constant averaged pair-distribution function}
\begin{figure}
\includegraphics{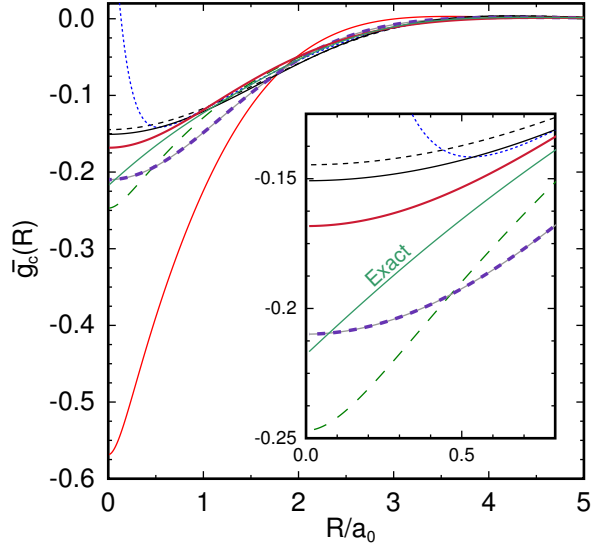}
\caption{\label{fig.gbar}
Coupling-constant averaged pair-distribution function $\bar{g_c}(R)$
calculated at $r_s = 2$ for the model HEG XC-kernels (see Fig.~\ref{fig.kernels}
for color code), compared to the ``exact'' parameterization
given in equation (36) of Ref.~\citenum{Perdew19922} 
(see also Ref.~\citenum{Perdew1997}).
The inset shows the zoomed region around $\bar{g_c}(R=0)$.
}
\end{figure}
Clearly all of the considered kernels greatly improve
the correlation energy of the HEG compared to the RPA.
The common characteristics shared by the kernels are that
they satisfy the exact $k \rightarrow 0$, $\w =0$ limit of the HEG
(except the rALDA, which neglects the correlation contribution
in equation~\ref{eq.A}), and that they decay for wavevectors above $2k_F$.
This decay is essential to an accurate description of the energetics of the HEG,
with the ALDA (which does not decay at large $k$),
yielding a correlation energy which is too positive.\cite{Olsen2012}
However the fact that we only observe small variations between 
the kernels considered in Fig.~\ref{fig.kernels}(d) indicates that 
the precise form of this decay is less important.

It is however interesting to consider the coupling-constant
averaged pair-distribution function $\bar{g_c}(R)$,\cite{Perdew19922}
obtained as the Fourier transform of
$\bar\varepsilon_c(k)$ multiplied by
$\pi/[2nk_F]$.\cite{Lein2000}
The pair-distribution function $g_c(R)$ is obtained
from the derivative of $\bar{g_c}(R)$ with 
respect to $r_s$,
and the exact $g_c$ and $\bar{g_c}$ both satisfy cusp conditions
such that their slopes at $R=0$ are generally 
nonzero.\cite{Kimball1973, Perdew19922}
Concentrating on $\bar{g_c}(R)$, we note that in order to describe
a cusp in real space we require Fourier components
[i.e.\ nonzero $\bar\varepsilon_c(k)$] at large $k$.
Indeed the analysis of Ref.~\citenum{Kimball1973} finds that a frequency-independent
kernel must decay as $-\gamma/k^2$, where $\gamma  < 4\pi$, i.e.\ the Hartree-XC kernel 
retains a $1/k^2$ term at large $k$.
By considering Fig.~\ref{fig.kernels}(a) we see that the rALDA, 
CP and CPd kernels all decay as $-4\pi/k^2$ such that their Hartree-XC
kernels vanish, so that $\bar\varepsilon_c(k)$ also quickly
tends to zero at large wavevectors [Fig.~\ref{fig.kernels}(c)].
Thus these kernels cannot describe the cusp.

To illustrate this behavior, in Fig.~\ref{fig.gbar} we plot
$\bar{g_c}(R)$ calculated at $r_s=2$ for the different kernels,
compared to the RPA and to the parameterization of 
Refs.~\citenum{Perdew19922,Perdew1997}.
It is clear that the coupling-constant averaged
pair-distribution functions calculated for the 
rALDA, CP and CPd kernels are far softer
than those calculated for the RPA and CDOPs, whose Hartree-XC
kernels decay $\propto 1/k^2$.
Meanwhile as noted above the local $C$ term of the CDOP 
kernel causes a divergence in $\bar{g_c}(R)$.

The slope of coupling-constant averaged pair-distribution 
function calculated in the RPA 
is too steep, while it is improved for CDOPs.
In Ref.~\citenum{Lu2014} it was also found that $g_c(R)$ calculated
for CDOPs agreed well with the exact result.
The good performance of the CDOPs kernel for calculations of 
the pair-distribution
function might have been anticipated from the fact that the coefficient
$B$ appearing in equation~\ref{eq.cdop} itself is determined 
from $g(R=0)$.\cite{Moroni1995}
The above analysis illustrates how  the precise 
large-$k$ behavior of a kernel affects its description 
of the cusp of $g_c(R)$,
despite playing a lesser
role in the calculation of energetics.
\subsection{Applying HEG kernels to inhomogeneous systems}
\label{sec.averaging}

In order to calculate the correlation energy of an inhomogeneous
system through equation~\ref{eq.Ec} we require
$\fxc$ evaluated in a plane-wave basis,
which in general is constructed from the real-space 
kernel through
\begin{equation}
\fxc^{\vec{G}\vec{G'}}(\vec{q},\w)=
\frac{1}{V}
\int_V d\vec{r}
\int_V d\vec{r'}
e^{-i(\vec{q} + \vec{G})\cdot\vec{r}}
\fxc(\vec{r},\vec{r'},\w) 
e^{i(\vec{q} + \vec{G'})\cdot\vec{r'}}
\label{eq.FT}
\end{equation}
where $V$ is the volume of the entire crystal,
consisting of $N_q$ replicas of the unit cell of volume $\Omega$.
The question is how to incorporate into this formalism 
a model ($m$) kernel which has the form $\fxc^m(n,k,\w)$ or
$\fxc^m(n,|\vec{r}-\vec{r'}|,\w)$.
In the case that the system is homogeneous ($n(\vec{r}) = n$), we
simply make the substitution $\fxc(\vec{r},\vec{r'},\w)  \rightarrow 
\fxc^m(n,|\vec{r} - \vec{r'}|,\w)$ to get a diagonal kernel,
\begin{equation}
\fxc^{\mathrm{hom} \ \vec{G}\vec{G'}}(\vec{q},\w) = \delta_{\vec{G}\vec{G'}} \fxc^m(n,|\vec{q}+\vec{G}|,\w).
\label{eq.homdens}
\end{equation}
Alternatively, if the model kernel is fully local (independent of $k$, e.g.\  the ALDA), it is 
natural to choose the local density to construct the kernel, and obtain
\begin{equation}
\fxc^{\mathrm{loc} \ \vec{G}\vec{G'}}(\vec{q},\w) = \frac{1}{\Omega} \int_\Omega d\vec{r} 
e^{-i(\vec{G}-\vec{G'})\cdot\vec{r}} \fxc^m(n(\vec{r}),\w).
\label{eq.localkernel}
\end{equation}
However for nonlocal kernels and inhomogeneous systems it is not obvious how
one should construct $\fxc(\vec{r},\vec{r'},\w)$, except for two
requirements.
First, an arbitrary model kernel should be symmetric in
$\vec{r}$ and $\vec{r'}$:\cite{Burkebook}
\begin{equation}
\fxc(\vec{r},\vec{r'},\w) = \fxc(\vec{r'},\vec{r},\w).
\label{eq.rsymm}
\end{equation}
Second, for the JGMs kernel, we require that in the
$q\rightarrow 0 $ limit the head of $\fxc$ (i.e.\ $\vec{G} = \vec{G'} = 0)$
diverges as $1/q^2$ while the wings 
($\vec{G} \neq \vec{G'}$ = 0) diverge no faster than $1/q$ (Ref.~\citenum{Ghosez1997}).
As shown below, this second requirement turns out to exclude previous schemes 
used in ACFD-DFT calculations, which we now briefly review.

\subsubsection{Density symmetrization}

A symmetric kernel can be obtained by making the following substitution
into equation~\ref{eq.FT}:
\begin{equation}
\fxc(\vec{r},\vec{r'},\w)  \rightarrow \fxc^m(\mathcal{S}[n],|\vec{r}-\vec{r'}|,\w).
\label{eq.denssym}
\end{equation}
Here, $\mathcal{S}$ is a functional of the density symmetric in $\vec{r}$ and $\vec{r'}$,
whose possible forms span a wide range of complexity.\cite{Garcia1996, Cuevas2012}
Refs.~\citenum{Jung2004}~and~\citenum{Olsen2013}
take a two-point average, $\mathcal{S}[n] = 1/2[n(\vec{r}) + n(\vec{r'})]$,
which has an intuitive interpretation.

A disadvantage of the symmetrization in equation~\ref{eq.denssym} is that
in order to evaluate equation~\ref{eq.FT}, it is necessary to work 
with a real-space representation
of the kernel and perform an integration over the entire volume of the 
crystal.
For short-range kernels this integral can be converged by sampling over 
a number of unit cells,\cite{Olsen2013} but for a long-range kernel like 
$\fxcJGMs$ [decaying as $-\a/(4\pi R)$] the required sampling might be 
prohibitively large.
Also, since the kernel is constructed in real space, it is not obvious how
to control the $1/q$ divergences of the JGMs kernel in reciprocal space.
Finally we note that the $1/R$ real-space divergence of the kernels
leads to slow convergence with the real-space grid used to evaluate
the integral in equation~\ref{eq.FT}.\cite{Olsen2013}

\subsubsection{Kernel symmetrization}
An alternative approach followed in Ref.~\citenum{Lu2014} is  to start 
from a nonsymmetric form of $\fxc$, which we label $\fxc^\mathrm{NS}$:
\begin{equation}
\fxc^\mathrm{NS}(\vec{r},\vec{r'},\w) = \fxc^m\left(n(\vec{r}),|\vec{r}-\vec{r'}|,\w\right).
\end{equation}
Inserting $\fxc^\mathrm{NS}$ into equation~\ref{eq.FT} gives
\begin{equation}
\fxc^{\mathrm{NS},\vec{G}\vec{G'}}(\vec{q},\w) = 
\frac{1}{\Omega} \int_\Omega d\vec{r} e^{-i(\vec{G} 
- \vec{G'})\cdot\vec{r}} \fxc^m\left(n(\vec{r}),|\vec{q}+\vec{G'}|,\w\right).
\label{eq.nonsymkern}
\end{equation}
A symmetric kernel 
can then be obtained by averaging 
$\fxc^{\mathrm{NS},\vec{G}\vec{G'}}$ with its Hermitian conjugate,
i.e.\
\begin{equation}
\fxc^{\mathrm{S},\vec{G}\vec{G'}}(\vec{q},\w) = \frac{1}{2} \left(\fxc^{\mathrm{NS},\vec{G}\vec{G'}}(\vec{q},\w)
+
\left[\fxc^{\mathrm{NS},\vec{G'}\vec{G}}(\vec{q},\w)\right]^*\right).
\label{eq.Lu_av}
\end{equation}
This procedure can be seen equivalently\cite{Lu2014} as inserting the symmetric
combination $1/2 [\fxc^\mathrm{NS}(\vec{r},\vec{r'},\w)  + \fxc^\mathrm{NS}(\vec{r'},\vec{r},\w) ]$
into equation~\ref{eq.FT}, and therefore corresponds to a two-point average
of the kernel; in the case that the kernel is linear in density,
this scheme is equivalent to averaging the density.
Equation~\ref{eq.nonsymkern} has the computational advantages that the integral
is over a single unit cell and requires only that the density (and not the kernel)
is represented on the real-space grid.
However, considering the JGMs XC-kernel, upon 
inserting equation~\ref{eq.JGMs} into equation~\ref{eq.nonsymkern}
and performing the average of equation~\ref{eq.Lu_av}
we are left with a matrix whose wings 
diverge $\propto1/q^2$ as $q\rightarrow0$, not $1/q$ as required.
Therefore equation~\ref{eq.Lu_av} is unsuitable for the current study.

\subsubsection{Wavevector symmetrization}
In order to correctly deal with the JGMs kernel while
retaining some of the computational advantages of 
equation~\ref{eq.nonsymkern},
we follow the approach of Ref.~\citenum{Trevisanutto2013}
and symmetrize the wavevector appearing in the right hand
side of equation~\ref{eq.nonsymkern} with the substitution
$|\vec{q} + \vec{G'}| \rightarrow \sqrt{|\vec{q}+\vec{G}||\vec{q}+\vec{G'}|}$:
\begin{equation}
\fxc^{\vec{G}\vec{G'}}(\vec{q},\w) = \frac{1}{\Omega} \int_\Omega  d\vec{r} e^{-i(\vec{G} - \vec{G'})\cdot\vec{r}}  
\fxc^m\left(n(\vec{r}),\sqrt{|\vec{q}+\vec{G}||\vec{q}+\vec{G'}|},\w\right).
\label{eq.kernel_FT}
\end{equation}
Like equation~\ref{eq.nonsymkern}, equation~\ref{eq.kernel_FT} requires
the integral over the unit cell only and deals with the reciprocal-space
form of the kernels.
Using equation~\ref{eq.kernel_FT} to construct the JGMs kernel yields
a matrix whose head and wings diverge in the  $q \rightarrow 0 $ limit 
as $1/q^2$ or $1/q$, as required, while the hermiticity of
$ \fxc^{\vec{G}\vec{G'}}(\vec{q},\w)$ automatically satisfies
the symmetry requirement (equation~\ref{eq.rsymm}).
Trivially, the averaging scheme will reproduce the equations~\ref{eq.homdens}
and~\ref{eq.localkernel} when applied to systems with a homogeneous
density or a local kernel, and
furthermore the diagonal ($\vec{G}=\vec{G'}$) elements coincide
with those calculated with the two-point average of the kernel,
equation~\ref{eq.Lu_av}.

Since this work is concerned with the comparison of a large number
of kernels, we have elected to use equation~\ref{eq.kernel_FT} 
on the grounds that it is relatively efficient, and can deal
with the divergences in the JGMs kernel correctly.
However the physical interpretation of the
off-diagonal elements arising from the wavevector-symmetrization 
is not transparent.
Although a two-point scheme also suffers from limitations
(e.g.\ the two-point kernel has no
knowledge of the medium lying between $\vec{r}$ and $\vec{r'}$),
it still remains a more intuitive procedure.
The fact that we have to invoke an averaging scheme at all is
an undesirable consequence of using HEG kernels to describe 
inhomogeneous systems.
In reality the use of different schemes can only be 
justified through testing and comparison
with experiments or other calculations, such as that 
performed in Refs.~\citenum{Olsen2012,Olsen2013,Olsen2014,Lu2014}
and here.

\subsection{Computational details}
\label{sec.comp_details}

All calculations in this work were performed using the \texttt{GPAW} 
code.\cite{Enkovaara2010}
The Kohn-Sham states and energies used to construct the response
function (equation~\ref{eq.chiPW}) were calculated using the local-density
approximation to DFT\cite{Hohenberg1964, Kohn1965,Perdew1992} 
within the projector-augmented wave (PAW) framework.\cite{Blochl1994}
We used $6\times6\times6$ and $12\times12\times12$ unshifted
Monkhorst-Pack\cite{Monkhorst1976}
meshes to sample the Brillouin zone for insulators and metals
respectively, and constructed the occupation factors for each Kohn-Sham 
state using  a Fermi-Dirac distribution function of width 0.01~eV.
For H, He and H$_2$ we used a simulation cell of 6$\times$6$\times$7 \AA$^3$ and
$\Gamma$-point sampling.

When calculating $E_\mathrm{c}$ the wavefunctions were expanded in a plane-wave basis set
up to a maximum kinetic energy of 600~eV.
Following previous studies,\cite{Harl2010,Olsen2013} we used the frozen-core
approximation but included semicore
states for some elements.\cite{PAWnote}
We note that norm-conservation was not enforced in the generation of
our PAW potentials, 
while it is reported in Ref.~\citenum{Klimes20142}
that including norm-conservation might increase the magnitude of the RPA correlation
energy and decrease the calculated lattice constants for certain materials.
As a result some care should be taken in making comparisons to experiment,
although  we expect calculations including the XC-kernel
to be less sensitive to this convergence issue (section~\ref{sec.rangeRPA}).

For the matrices representing the response function and kernel,
we used a lower plane-wave cutoff $E_\mathrm{cut}$ of
400~eV (300~eV for Na and H$_2$), and a $\vec{q}$ grid matching
the Brillouin zone sampling of the ground-state calculation.
We truncated the sum-over-states appearing in equation~\ref{eq.chiPW}
at a number of bands equal to the number of plane waves describing
the response function, e.g.\ $\sim$700 for Si.
Within this approximation, the following extrapolation scheme is
commonly used for the RPA correlation energy:\cite{Harl2010,Klimes20142}
\begin{equation}
E^\mathrm{RPA}_\mathrm{c}(E_\mathrm{cut}) \sim E^\mathrm{RPA}_\mathrm{c}(E_\mathrm{cut}\rightarrow\infty)
+ K E_\mathrm{cut}^{-3/2}.
\label{eq.extrap}
\end{equation}

\begin{figure}
\includegraphics{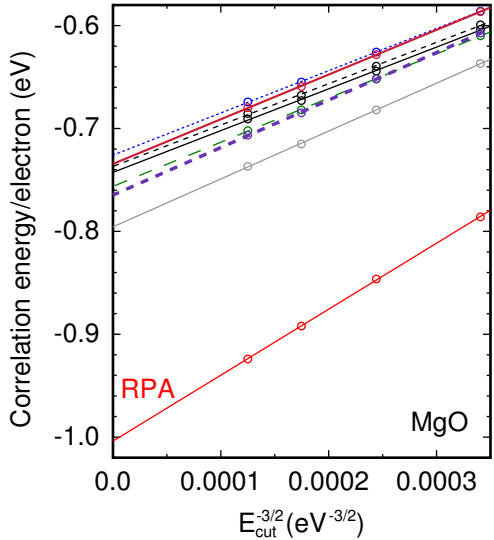}
\caption{\label{fig.MgO}
The correlation energy $E_\mathrm{c}$ evaluated per electron for
MgO at a lattice constant of 4.23~\AA, using different
approximations for the XC-kernel (see Fig.~\ref{fig.kernels}
for the color code).
$E_\mathrm{cut}$ is the plane-wave cutoff used for
the matrices representing the response function and  XC-kernel.
The circles represent calculated data points, and the
lines are fits from equation~\ref{eq.extrap}.
}
\end{figure}

In Ref.~\citenum{Olsen2013} it was proposed that the same
expression can be applied to the correlation energy
calculated with the rALDA kernel.
We have tested this expression for each of the kernels
in section~\ref{sec.kernelzoo} for a set of 10 materials
(see section~\ref{sec.lattice}).
As an example, in Fig.~\ref{fig.MgO} we plot the correlation 
energy per electron
calculated for MgO as a function of $E_\mathrm{cut}^{-3/2}$.
As demonstrated by the straight lines, equation~\ref{eq.extrap}
apparently gives a good description of the correlation energy
calculated for $E_\mathrm{cut}>$200~eV for all of the kernels.
We have observed the same behavior across the combinations of materials
and kernels.
Therefore in order to facilitate comparison across the entire test set
we will apply equation~\ref{eq.extrap} for all XC-kernels.
We point out that the correlation energy tends to converge faster
(shallower lines in Fig.~\ref{fig.MgO}) when a nonzero $\fxc$ is used, and
for calculating structural properties the extrapolation
is often unnecessary.

Constructing the XC-kernel with equation~\ref{eq.kernel_FT}
is not straightforward
due to the dual dependence on $\vec{G}$ and $\vec{G'}$.
Our current implementation distributes the rows of the $\fxc$
matrix among processors before evaluating the integral in
equation~\ref{eq.kernel_FT}.
In the future it may be appropriate to improve performance 
through an interpolation scheme, as for Refs.~\citenum{Lu2014}~and~\citenum{Roman2009}.
On the other hand, for the systems studied here the time taken to 
construct the kernel is small compared to
that spent constructing the response function $\chi_\mathrm{KS}$ and inverting
equation~\ref{eq.chi_ks}.
When constructing the kernel, we use the PAW all-electron density
to be consistent with previous work.\cite{Olsen2013}
The $1/q^2$ divergence of the Coulomb interaction and
JGMs kernel was treated within the scheme described in Ref.~\citenum{Yan2011}.

Having obtained $\chi^\l$ we evaluated equation~\ref{eq.Ec}
by numerical (Gauss-Legendre) integration over the coupling
constant ($N_\l = 8$ points) and frequency ($N_\w = 16$ points, using a logarithmic
mesh).\cite{Olsen20132}
By virtue of the scaling relation (equation~\ref{eq.scaling})
we must construct the rALDA kernel once, a general static kernel 
$N_\l$ times and a dynamical kernel $N_\l N_\w$ times; hence there
is a prefactor of $\sim 100$ applied to computing $\fxcCPd$ compared
to $\fxcrALDA$.

To calculate structural properties, we
evaluated the total energy $(E_\mathrm{Tot}^\mathrm{LDA} - E_{xc}^\mathrm{LDA})
+E_\mathrm{x} + E_\mathrm{c}$ for seven lattice constants centered around the experimental
value and fit the values to the Birch-Murnaghan equation of state.\cite{Birch1947}
We used higher plane-wave cutoffs of 800~eV (900 eV for MgO, LiCl and LiF)
to evaluate the LDA energies and $E_\mathrm{x}$, and used a denser
sampling of the Brillouin zone combined with the 
Wigner-Seitz truncation scheme described in Ref.~\citenum{Sundararaman2013}
to calculate $E_\mathrm{x}$.
We typically obtain converged exchange energies for insulators with
a sampling of $10\times10\times10$, while metals 
require a denser sampling\cite{Sundararaman2013}
(e.g.\ $20\times20\times20$).
Since the bulk modulus is constructed from derivatives
of the energy, it is rather prone to numerical error, 
to the extent that different code
implementations of the same method can yield different 
results.\cite{Labat2013}
In this respect one should attach more significance to the
calculated lattice constants than bulk moduli, since
the former are more robust quantities.
However even for the bulk moduli one expects a 
reduction in error when
comparing different XC-kernels
within the same computational framework.

\section{Results and discussion}
\label{sec.results}

\begin{figure*}
\includegraphics[width=175mm]{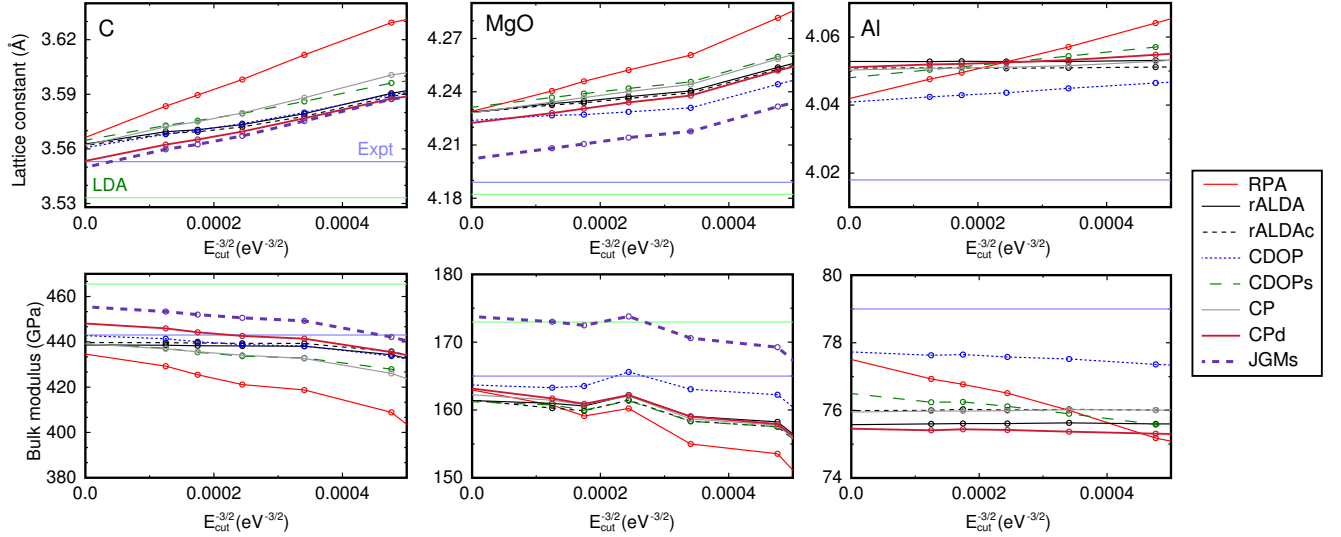}
\caption{\label{fig.materials}
Lattice constants and bulk moduli calculated with different 
XC-kernels for
C, MgO and Al,
vs $E_\mathrm{cut}^{-3/2}$, where
$E_\mathrm{cut}$ is 
the plane-wave cutoff of the response function and XC-kernel
matrices.
The green and blue horizontal lines give the values calculated
with the LDA and obtained from experiment, respectively, where
the experimental data were tabulated in Ref.~\citenum{Harl2010}
(note the LDA lattice constant/bulk modulus for Al [3.99~\AA/83~GPa]
is off the scale).
The values at infinite cutoff ($E_\mathrm{cut}^{-3/2} = 0$) 
were calculated from the correlation energies
extrapolated from  equation~\ref{eq.extrap}.
}
\end{figure*}

\subsection{Lattice constants and bulk moduli}
\label{sec.lattice}

We have selected a test set of 10 materials,
consisting of 3 tetrahedrally-bonded semiconductors
(diamond C, Si and SiC), 3 ionic compounds (MgO, LiCl
and LiF) and 4 metals (Al, Na, Cu and Pd).
For each material, we used the XC-kernels introduced
in section~\ref{sec.kernelzoo} to calculate the
lattice constant and bulk modulus.
Here we compare these results to those obtained from
DFT (in the LDA or from the generalized-gradient PBE functional\cite{Perdew1996}), 
the RPA, and to the experimental
values tabulated in Ref.~\citenum{Harl2010}.

\subsubsection{General trends}

Figure~\ref{fig.materials} shows the lattice constants
and bulk moduli calculated for C, MgO and Al as a function
of $E_\mathrm{cut}^{-3/2}$, a quantity inversely proportional to 
the number of plane-waves describing
the response function $\chi_\mathrm{KS}$
and XC-kernel $\fxc$ (c.f.\ Fig.~\ref{fig.MgO}).
The quantities at $E_\mathrm{cut}^{-3/2} = 0$ 
were calculated from $E_\mathrm{c}$ extrapolated to 
infinite $E_\mathrm{cut}$ using equation~\ref{eq.extrap}.
We also show the values obtained from the LDA and experiment as horizontal lines.

There are three key observations to be made from
Fig.~\ref{fig.materials}.
First, for non-metallic systems the rALDA, rALDAc, CDOPs and CP
kernels yield almost identical results, which in turn
are very similar to the RPA.
Second, the JGMs, CPd and the CDOP kernels (which respectively
are long-range, dynamical or have a local term) display
distinct behavior. 
For instance the JGMs kernel predicts smaller lattice constants
and larger bulk moduli than the other XC-kernels.
Finally, all of the XC-kernels show faster convergence
with respect to $E_\mathrm{cut}$ compared to the RPA,
as found for the correlation energy (Fig.~\ref{fig.MgO}).

\begin{figure*}
\includegraphics[width=175mm]{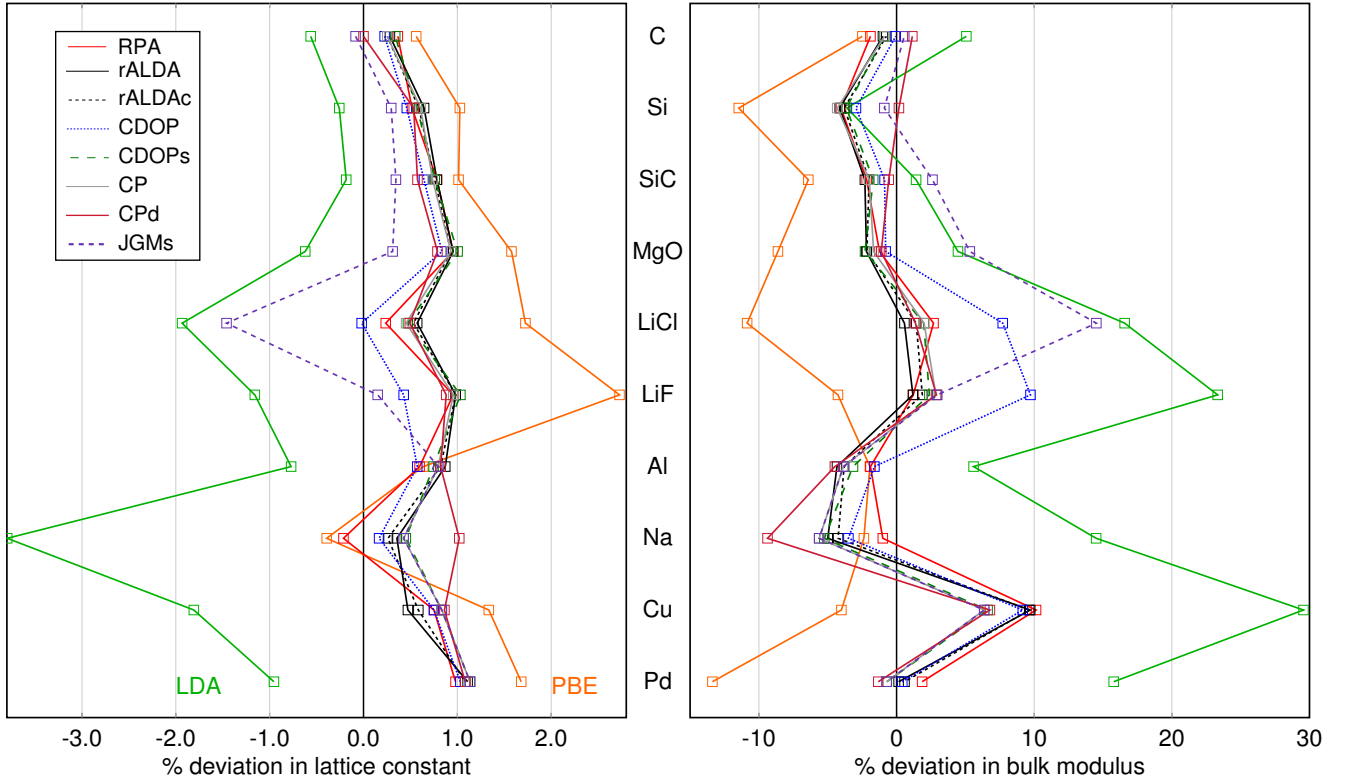}
\caption{\label{fig.latt_bm}
Percentage deviation from experiment of calculated 
lattice constants and bulk
moduli for the test set of 10 materials.
The values used to construct the plots are presented in Table~\ref{tab.latt_bm}.
Each line corresponds to a different approximation for $\fxc$.
}
\end{figure*}

Keeping the above points in mind, we 
extend this analysis to the full test set and
consider each kernel in turn.
The entire dataset is given in Fig.~\ref{fig.latt_bm} and Table~\ref{tab.latt_bm}.

\begin{table*}
\caption{\scriptsize
\label{tab.latt_bm} 
Lattice constants (in \AA) and bulk moduli (GPa) calculated for the test set of 10 materials
compared to the experimental data tabulated in Ref.~\citenum{Harl2010}.
The results were calculated from correlation energies obtained by the extrapolation
procedure of equation~\ref{eq.extrap}.
The mean absolute error (M.A.E.) compared to experiment is shown in the final
row.
The experimental lattice constants were corrected for expansion due to zero-point
motion; the bulk moduli have not been corrected.
For comparison the LDA and RPA calculations of Ref.~\citenum{Harl2010} are also
presented (note that these RPA calculations were performed on top of Kohn-Sham
states obtained within the generalized-gradient approximation\cite{Perdew1996}).
The CP and JGMs kernels coincide for metallic systems.
}
\begin{tabular}{lccccccccccccc} 
\hline
\hline 
      &   LDA  & LDA\footnotemark[1] &  PBE   &RPA&  RPA\footnotemark[1] &  rALDA    & rALDAc    &  CDOP &  CDOPs  &  CP  &  CPd  &  JGMs   & Exp. \\
\hline
C     & 3.533  &3.534&3.573&3.566&3.572 &     3.563 &     3.562 & 3.561 & 3.565   &3.562 & 3.553 &  3.550  & 3.553\\
Si    & 5.407  &5.404&5.477&5.449&5.432 &     5.456 &     5.453 & 5.446 & 5.452   &5.454 & 5.450 &  5.437  & 5.421\\
SiC   & 4.338  &4.332&4.390&4.380&4.365 &     4.380 &     4.379 & 4.374 & 4.379   &4.378 & 4.371 &  4.361  & 4.346\\
MgO   & 4.163  &4.169&4.255&4.229&4.225 &     4.229 &     4.228 & 4.224 & 4.231   &4.228 & 4.222 &  4.202  & 4.189\\
LiCl  & 4.972  &4.967&5.157&5.082&5.074 &     5.099 &     5.097 & 5.069 & 5.094   &5.093 & 5.095 &  4.996  & 5.070\\
LiF   & 3.926  &3.913&4.080&4.010&3.998 &     4.011 &     4.011 & 3.989 & 4.013   &4.010 & 4.007 &  3.978  & 3.972\\
Al    & 3.987  &3.983&4.044&4.042&4.037 &     4.053 &     4.051 & 4.041 & 4.048   &4.050 & 4.051 &  4.050  & 4.018\\
Na    & 4.054  &4.056&4.197&4.205&4.182 &     4.229 &     4.225 & 4.221 & 4.233   &4.232 & 4.257 &  4.232  & 4.214\\
Cu    & 3.530  &3.523&3.643&3.622&3.597 &     3.612 &     3.616 & 3.622 & 3.625   &3.625 & 3.626 &  3.625  & 3.595\\
Pd    & 3.839  &3.830&3.941&3.914&3.896 &     3.919 &     3.918 & 3.916 & 3.920   &3.920 & 3.918 &  3.920  & 3.876\\
\hline
\% M.A.E.& 1.2 &1.3  &1.3  &0.6  &0.5   &        0.7&     0.7   & 0.5   & 0.7     & 0.7  &   0.7 &   0.6   & --- \\
\hline
\hline
\hspace{2 cm}
\end{tabular}

\begin{tabular}{lccccccccccccc}
\hline
\hline
      &   LDA  & LDA\footnotemark[1] & PBE  &  RPA&  RPA\footnotemark[1] &  rALDA    & rALDAc    &  CDOP &  CDOPs  &  CP  &  CPd  &  JGMs   & Exp. \\
\hline
C      &  465   &  465   &  432 &435 &441   &     439 &     440 &  443   &  439    & 439  & 448   &  455    & 443  \\
Si     &   95   &  97    &   88 & 95 & 99   &     95  &     95  &   96   &   95    &  95  &  99   &   98    & 99   \\
SiC    &  228   &  229   &  211 &220 & 223  &     220 &     221 &  223   &  221    & 221  & 224   &  231    & 225  \\
MgO    &  172   &  172   &  151 &163   & 168  &161      &  161    &  164   &  161    & 162  & 163   &  174    & 165  \\
LiCl   &   41   &   41   &   31 &36    & 37   &    35   &   35    &   38   &   36    &  36  &  35   &   40    & 35   \\
LiF    &   86   &   87   &   67 &71   & 76   &71       &   71    &   77   &   72    &  72  &  72   &   72    & 70   \\
Al     &   83   &   84   &   77 &78   & 77   &76       &      76 &   78   &   76    &  76  &  75   &   76    & 79   \\
Na     &    9   &    9   &    8 &8   &  8   &   8     &    8    &   8    &    8    &   8  &   7   &    8    & 8    \\
Cu     &  184   &  186   &  136 &156   & 153  &   156   &   156   &   155  &  151    & 151  & 152   &  151    & 142  \\
Pd     &  226   &  226   &  169 &199   & 181  &   195   &   196   &   196  &  194    & 194  & 192   &  194    & 195  \\
\hline
\% M.A.E.& 12   & 12     &    7 &3     & 3    &  3      &    3    & 4      & 3      &  3    &    3  &   4     & --- \\
\hline
\hline
\end{tabular}
\footnotetext[1]{Ref.~\citenum{Harl2010}}
\end{table*}

\subsubsection{LDA, PBE and RPA}
The LDA typically underestimates lattice constants and overestimates
bulk moduli, while PBE displays opposite behavior.
For tetrahedral semiconductors the LDA is difficult to beat,
and is by far the most computationally-efficient scheme.
Using exact exchange and the RPA correlation energy  yields
improved bulk moduli and lattice constants (e.g.\ a mean absolute error
in lattice constants of 0.6\% compared to 1.2\% for the LDA).
Apart from Na, the calculated RPA lattice constants are larger than
the experimental values, a result also found in Ref.~\citenum{Harl2010}.

\subsubsection{rALDA and rALDAc}

For the non-metallic systems, the rALDA and rALDAc kernels
produce lattice constants and bulk moduli which are essentially indistinguishable
from each other.
In turn, these results are in close agreement with the RPA.
For metals, one can identify differences between the kernels, although
the magnitude of variation is still very small ($<$0.1\%).
The close agreement between rALDA and rALDAc confirms that the exchange 
contribution dominates in equation~\ref{eq.A}, and supports the use
of the exchange-only rALDA kernel.

The rALDA kernels also display the fastest
convergence with respect to $E_\mathrm{cut}$.
Recalling the form of the kernels (equation~\ref{eq.ralda}),
the components of $\fHxc$ are truncated
for wavevectors exceeding the cutoff $k_c$.
For a homogeneous system (equation~\ref{eq.chi_HEG}),
for $k'>k_c$, the interacting and non-interacting response functions coincide
and therefore the contribution to the correlation energy
at these wavevectors vanishes.
In inhomogeneous systems, high-density regions (large $k_c$) give
terms that converge like the RPA, 
but the rALDA convergence is still superior after the kernel
averaging procedure (equation~\ref{eq.kernel_FT}) is applied.

From this convergence behavior, we conclude that the
short-range description of correlation obtained from an 
XC-kernel like the rALDA is easier to describe in 
a plane-wave basis than the erroneous short-range behavior of the RPA.
This result might have been anticipated from the HEG, where 
 the coupling-constant averaged
pair-distribution function calculated for the rALDA is 
softer than for the  RPA (Fig.~\ref{fig.gbar}).

\subsubsection{CP and CDOPs}

The CP and CDOPs kernels yield lattice constants and bulk
moduli which are also very similar to each other 
across the full range of systems.
This behavior can be explained by considering Fig.~\ref{fig.kernels}(a),
where it can be seen that $\fxcCDOPs$ lies more negative than  $\fxcCP$ for $k$ less
that $\sim2k_F$, and more positive otherwise.
The kernel averaging procedure smears out these differences.
In particular, there is a negligible effect from modifying
the large-$k$ behavior from $-4\pi B/ k^2$ (CDOPs) to $-4\pi / k^2$ (CP).

$\fxcCDOPs$ and $\fxcCP$ closely follow the rALDA kernels (and the RPA) 
for non-metallic systems.
For metallic systems, differences of order 0.3\% can be observed.
The most likely reason for this difference is the long-range behavior
of the rALDA kernels, which display decaying oscillations, compared
to the CDOPs and CP kernels which go to zero more smoothly.
The small positive hump displayed by the CDOPs kernel in real space [Fig.~\ref{fig.kernels}(b)]
appears to have little effect on the correlation energy.

\subsubsection{CDOP}
\label{sec.cdop_results}

The CDOP kernel (equation~\ref{eq.cdop}) differs 
from $\fxcCDOPs$ by having a local term.
This local term has a noticeable effect on the calculated
structural properties, with the CDOP kernel predicting
slightly smaller and larger lattice constants and bulk
moduli, respectively.
Indeed the CDOP kernel displays the closest agreement
with experimental lattice constants, but performs less
well on bulk moduli.

In section~\ref{sec.paircorrelation} it was pointed out that
the local term in the CDOP kernel
leads to a divergent pair-distribution function.
The local term may also be expected to introduce convergence
problems, as demonstrated for the (entirely local) ALDA kernel.\cite{Furche2005,Olsen2012}
In the current work, we have not found any significant
difference in the convergence behavior of the CDOP and CDOPs kernels when
calculating lattice constants and bulk moduli
for $E_\mathrm{cut}\leq$ 400~eV.
Only in cases where the RPA correlation energy converges relatively 
quickly (e.g.\ Al) can we observe a slowly-converging positive contribution 
to the CDOP correlation energy which is reminiscent of that found for the ALDA, 
c.f.\ Fig.~3 of Ref.~\citenum{Olsen2012}.
However, unlike for the ALDA, the magnitude of this contribution is very small
compared to the RPA-like convergence (e.g.\ Fig.~\ref{fig.MgO}).

\subsubsection{CPd}

The dynamical CPd kernel displays slight differences
to its $\w=0$ limit, $\fxcCP$.
Compared to the static kernels where the range of $\fxc$
is fixed by the density, for the CPd kernel the frequency appearing
in the denominator of $\chi_\mathrm{KS}$
also affects the range.
Interestingly the CPd bulk moduli of insulators
are slightly closer to experiment.
In other cases we find that
the CP and CPd kernels predict similar
results except for Na, where the CPd kernel finds 
a larger lattice constant and smaller bulk modulus, 
and C, where the CPd lattice constant lies on top
of the experimental value.

The CPd results 
show that even a simple dynamical kernel can predict 
different structural properties.
This result is not obvious from studies on the
HEG, where tests on the more complicated frequency-dependent kernel of
Ref.~\citenum{Richardson1994} found dynamical effects to be
less important than nonlocality when calculating correlation energies.\cite{Lein2000}
Although we do not observe systematic improvement with
the CPd kernel, it would be interesting to investigate its performance
for systems with a greater degree of inhomogeneity, e.g.\ molecules and surfaces.

\subsubsection{JGMs}

The lattice constants calculated with the JGMs kernel
for insulating systems display the closest agreement 
with experiment out of all of the considered kernels,
except for the notable example of LiCl, where the JGMs lattice
constant is underestimated by 1.4\%.
However the agreement with experimental bulk moduli is poorer, 
in some cases (SiC, MgO) worse than the LDA.
For metallic systems, the JGMs and CP kernels coincide.

\begin{table}
\caption{
\label{tab.jgm} 
Parameters relating to the JGMs kernel.
The values of $\<\alpha\>$ were obtained by inserting
the experimental band gaps $E_g$ into equation~\ref{eq.alpha}, while
inserting the ``effective gaps'' $E_g^\mathrm{eff}$ 
yields the $\alpha_{\mathrm{LRC}}$ values 
reported in Ref~\citenum{Botti2004}
(these calculations were performed at the experimental
lattice constant).
The $\alpha_{\mathrm{LRC}}$ values for LiCl and LiF were obtained
from equation~(4) of Ref.~\citenum{Botti2004} using the dielectric
constants tabulated in Ref~\citenum{Ashcroftbook}.
The experimental (direct) band gaps were obtained from
Refs.~\citenum{Roberts1967,Zucca1970,Choyke1969,Whited1973,
Brown1970,Piacentini1976}.}
\begin{tabular}{lllllll}
\hline
\hline
                          & C    & Si   & SiC  & MgO  & LiCl & LiF   \\
\hline
$E_g$ (eV)                & 7.3  & 3.4  & 6.0  & 7.8  & 9.4  & 14.2  \\
$\<\alpha\>$              & 0.58 & 0.89 & 1.30 & 2.32 & 5.60 &  7.03 \\
\hline
$E_g^\mathrm{eff}$ (eV)   & 7.43 & 1.57 & 3.62 & 6.74 & 3.98 &  6.07 \\
$\alpha_{\mathrm{LRC}}$   & 0.6  & 0.2  & 0.5  & 1.8  & 1.5  &  2.2  \\
\hline
\hline
\end{tabular}
\end{table}

It is important to establish the importance of the value
of $E_g$.
In the current work we have used the experimental, direct gap (Table~\ref{tab.jgm}),
but we equally could have chosen the indirect gap, or even
defined a more general $\vec{r}$-dependent gap function.\cite{Fabiano2014}
An alternative option is to make the link 
to the description of excitons\cite{Trevisanutto2013,Botti2004}
and consider the head ($\vec{G} = \vec{G'} = 0)$  
of $\fxcJGMs$ in the $q\rightarrow 0$ limit,
which can be written as $-\<\alpha\>/q^2$ where
\begin{equation}
\<\alpha\> =  \frac{4\pi}{\Omega}\int_\Omega  d\vec{r} \
\left[ 1 -  e^{-E_g^2/[4\pi n(\vec{r})]} \right].
\label{eq.alpha}
\end{equation}
The values of $\<\alpha\>$ computed from equation~\ref{eq.alpha}
for the experimental gaps are given in Table~\ref{tab.jgm}.
These values can be compared to Ref.~\citenum{Botti2004},
where a long-range (LRC) attractive kernel was introduced
as $\fxc(R) = -\alpha_\mathrm{LRC}/(4\pi R)$.
We note that the head of this matrix in reciprocal space
in the $q\rightarrow0$ limit coincides with the JGMs
kernel with $\alpha_\mathrm{LRC} \rightarrow \<\alpha\>$,
and also that this single matrix element is considered
the most important for the calculation of excitonic 
effects.\cite{Botti2004}

From Table~\ref{tab.jgm} it is clear that the values
of $\<\a\>$ calculated with the experimental gaps and PAW
densities are somewhat larger than the values of $\alpha_\mathrm{LRC}$
reported in Ref.~\citenum{Botti2004}, which were found to give
a good description of excitonic effects in absorption spectra
of semiconductors and MgO.
To explore this point further we adopted an inverse approach and considered
an effective gap $E_g^\mathrm{eff}$, which when inserted
into equation~\ref{eq.alpha} yields $\alpha_\mathrm{LRC}$.
These LRC ``gaps'' are smaller than experimental values,
especially for the ionic compounds.
Indeed the empirical $\alpha_\mathrm{LRC}$ values of LiCl
and LiF are significantly smaller than those expected both
from the JGM or bootstrap kernels,\cite{Trevisanutto2013,Sharma2011}
which have been shown to accurately capture the exciton in LiF.

\begin{figure}
\includegraphics[width=50mm]{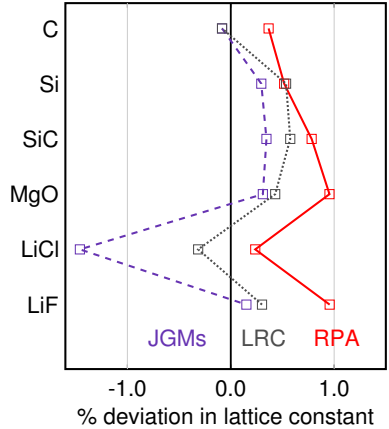}
\caption{\label{fig.LRC}
Percentage deviations of lattice constants compared to experiment,\cite{Harl2010}
calculated with the RPA, and the JGMs kernel
using the experimental direct band gaps or effective
LRC gaps (see Table~\ref{tab.jgm}).
}%
\end{figure}
We repeated the JGMs kernel calculations using the LRC gaps $E_g^\mathrm{eff}$,
and show the obtained lattice constants in Fig.~\ref{fig.LRC}.
The LRC results lie between the lattice constants calculated
with the RPA and with the JGMs kernel/experimental gaps,
and thus improve the LiCl result.
Comparison of LiCl and LiF demonstrates the nonlinear
relation between $E_g^\mathrm{eff}$ and the lattice constant.
In both cases the effective gap is reduced by more than 50\%
from its experimental value, but the effect on the LiCl lattice
constant is an order of magnitude larger than for LiF.

The improved agreement of lattice constants with experiment
compared to the RPA
shows that XC-kernels with long-range components
represent an interesting avenue to study.
A key question is whether the tendency for the JGMs kernel
to favor smaller lattice constants than the RPA 
is directly related to the fundamental long-range character of the
former, or
whether it is in fact a consequence of the precise
form of the kernel.
The strength of the long-range 
part of the JGMs Hartree-XC kernel is determined by 
$\exp(-E_g^2/4\pi n)$ (equation~\ref{eq.JGMs}), which
becomes RPA-like in the high density limit
and vanishes in the low density limit.
The RPA correlation energy is generally negative, while
a zero $\fHxc$ implies a zero correlation energy.
Interpolating these two limits implies that a more
negative (i.e.\ stable) JGMs correlation energy 
will correspond
to a higher density, thus favoring a lower lattice constant.
This observation also provides an explanation for the varying
behavior of the bulk modulus and also the strong nonlinearity
in the variation of the lattice constant with band gap,
since the energy-volume
relation is expected to be sensitive to the relative
magnitude of $E_g$ and $n$.

We note that the bootstrap approach\cite{Sharma2011} is
an alternative method of constructing a long-range kernel.
Since the bootstrap kernel is constructed from $\chi_\mathrm{KS}$,
using it would avoid both the input of $E_g$
and the averaging procedures
discussed in section~\ref{sec.averaging}.
However, it would be necessary
to ensure that the bootstrap kernel displayed
reasonable behavior in the large $k$-limit.

\subsection{Absolute correlation energies}
\label{sec.abs_corr}

\begin{figure*}
\includegraphics[width=160mm]{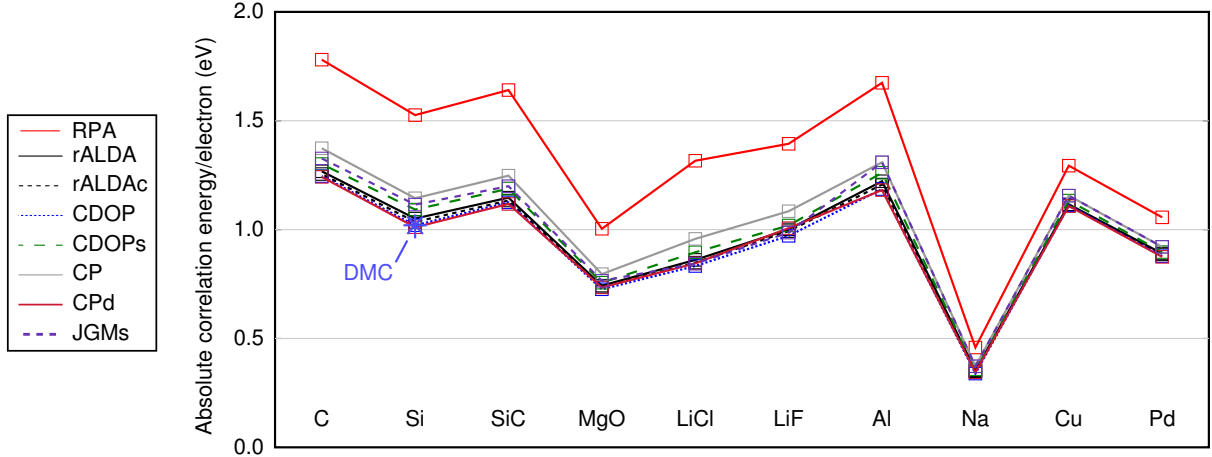}%
\caption{\label{fig.corr_energies}
Absolute correlation energies $E_\mathrm{c}$ (equation~\ref{eq.Ec})
calculated per (valence) electron for different kernels.
The correlation energy obtained for Si from the diffusion
Monte Carlo (DMC) calculations of Ref.~\citenum{Hood1998}
is also shown.
The core/valence partitioning of the PAW potentials
is given in section~\ref{sec.comp_details}.
The calculations were performed at the experimental lattice constant.
}
\end{figure*}

In Fig.~\ref{fig.corr_energies} we show the absolute correlation
energy per electron calculated using each of the different
kernels for the materials in the test set.
Absolute correlation energies are generally considered less
robust than properties constructed from energy differences,
being more difficult to converge and sensitive to details
of the PAW potentials.
However one can still perform a comparison between kernels,
and look for similarities with the trends observed
for the HEG [Fig.~\ref{fig.kernels}(d)].

The most obvious feature of Fig.~\ref{fig.corr_energies} is
the reduction of absolute correlation energy on moving from
the RPA to a nonzero $\fxc$, ranging from 0.1~eV for Na to 0.5~eV
for Si.
This change is the same order of magnitude as 
observed for the HEG.
The ordering of the HEG correlation energy calculated with different
kernels is also largely preserved, with the CP and CDOPs kernels
predicting larger magnitudes than CDOP and the rALDA kernels.

The difference between the rALDA and rALDAc kernels is small, with
the rALDA correlation energy being more negative by order
1\% or 0.01 eV per electron.
The difference between the static and dynamical forms of the CP kernel 
is an order of magnitude larger, with the static correlation energy 
being more negative.
Meanwhile the removal of the local term in the CDOP kernel
increases the magnitude of the correlation energy, with the CDOPs having a more negative
correlation energy than CDOP by 5\% or 0.05~eV per electron.

As in Ref.~\citenum{Lu2014} we can tentatively compare 
our calculated correlation energy for Si with the diffusion
Monte Carlo (DMC) calculations of Ref.~\citenum{Hood1998}.
Reassuringly the DMC correlation energy lies among the values
calculated with the model exchange kernels (Fig.~\ref{fig.corr_energies}), 
in fact displaying
closest agreement with CPd, rALDA and CDOP kernels.
We also note that our calculated CDOP correlation energy for Si (-1.02~eV per electron)  
lies on top of the value recently reported in Ref.~\citenum{Lu2014}
using a pseudopotential approximation and a different averaging scheme (equation~\ref{eq.Lu_av}).
With highly-accurate calculations of correlation energies in
extended systems now becoming a reality,\cite{Booth2013}
comparisons of this sort should become a useful test
for new kernels.

\subsection{Kernel averaging scheme}

It is interesting to compare the structural properties and 
correlation energies calculated using the symmetrized-wavevector
averaging scheme (equation~\ref{eq.kernel_FT})
to the two-point symmetrized density (equation~\ref{eq.denssym}).
The latter has previously been implemented for the rALDA,\cite{Olsen2013}
so here we restrict the comparison to this kernel.

Considering the lattice constants first,
we typically find a difference of 0.2\% 
between the two methods, with the symmetrized-density values larger 
than those calculated with the symmetrized-wavevector
in most cases.
Interestingly the agreement is worse for the bulk moduli, with an average
deviation of 6\%.
The absolute correlation energies also show a larger (4\%) deviation,
where using the two-point symmetrized density
scheme consistently yields more negative rALDA correlation energies than
the symmetrized-wavevector scheme, by an average of 0.04~eV per electron.

To understand the origin of these differences it is necessary
to consider the practical implementation of the two-point density
average (equation~\ref{eq.denssym}).
As mentioned in section~\ref{sec.averaging}, constructing the kernel
in this way involves sampling the $1/R$ Coulomb interaction
in real space.
The divergence at $R=0$ is replaced with a spherical average of $1/R$
taken over the volume per point in the real-space grid used to evaluate
the integral.~\cite{Olsen2013}
The absolute correlation energy is therefore rather sensitive to this grid spacing,
and its dependence on  volume (i.e.\ the bulk modulus) will also be 
difficult to converge.

The symmetrized-wavevector approach only samples the density
on the real space grid, and therefore shows a much weaker dependence on the
spacing between the grid points.
We verified this behavior for diamond C, where the
symmetrized-wavevector correlation energy changes by less than
$10^{-5}$ eV/electron on reducing the grid spacing from 0.17 to 0.11~\AA.
This is several orders of magnitude faster than the symmetrized-density
approach,\cite{Olsen2013} illustrating a computational advantage
of equation~\ref{eq.kernel_FT}.

\subsection{Spin and atomization energies: the H$\bf_2$ molecule}
\label{sec.H2}

\begin{table}
\caption{
\label{tab.H2} 
Correlation energies of H, H$_2$ and He and the atomization energy of the
H$_2$ molecule [$E_\mathrm{at}$(H$_2$)] calculated at
different levels of theory.  
The CCSD values are taken from
Ref.~\citenum{Lee2000} and the experimental atomization energy
from Ref.~\citenum{Karton2006}.
The rALDA results were calculated including spin-polarization,
with the kernels constructed either from equation~\ref{eq.kernel_FT}
or equation~\ref{eq.denssym}.
All values are given in eV.
}
\begin{tabular}{lrrrrr}
\hline
\hline
                        & RPA   & rALDA\footnotemark[1]& rALDA\footnotemark[2]& CCSD  & Exp   \\
\hline
H                       & -0.57 &       0.06           &         -0.02        & 0.00   & ---  \\
H$_2$                   & -2.22 &      -1.04           &         -1.22        & -1.11  & ---  \\
He                      & -1.82 &      -1.00           &         -1.08        & -1.14  & ---  \\
\hline
$E_\mathrm{at}$(H$_2$) &  4.74 &       4.82            &          4.85        &  4.75  & 4.75 \\
\hline
\hline
\footnotetext[1]{Symmetrized wavevector, equation~\ref{eq.kernel_FT}}
\footnotetext[2]{Symmetrized density, equation~\ref{eq.denssym}}
\end{tabular}
\end{table}

Throughout this study we have not considered any spin-dependence
of the XC-kernels.
However, the calculation of atomization or cohesive energies
usually requires the description of spin-polarized atoms
or molecules.
In this section we provide a demonstration of the importance
of spin by calculating 
the atomization energy of the H$_2$ molecule
with the rALDA kernel.

First, we note that the symmetrized-wavevector 
averaging procedure (equation~\ref{eq.kernel_FT})
can be equally applied to extended and finite systems.
In fact, for the rALDA kernel one can exploit the fact that
the Hartree-XC kernel strictly vanishes 
at any points in space where the density is less than 
$(|\vec{G}+\vec{q}||\vec{G'}+\vec{q}|)^{3/2}/(24\pi^2)$.
Therefore the Fourier transform can be performed in a 
small box which excludes the vacuum region generally required
to model isolated systems with periodic boundary conditions.
Since the H$_2$ molecule is spin-unpolarized we can calculate
its correlation energy without any further consideration, and
obtain a value of -1.04~eV with the rALDA kernel.
This value is within
0.1~eV of the value obtained from coupled-cluster calculations\cite{Lee2000}
and a significant improvement ($>$1~eV) over the RPA.
We find a similar level of agreement for the He atom (Table~\ref{tab.H2}).

For the spin-polarized H atom, following the analysis of Ref.~\citenum{Olsen2013}
we replace the integral equation (equation~\ref{eq.chi_ks}) with its
the spin-polarized version, valid for systems where only one spin channel is occupied:
\begin{equation}
\chi^{\uparrow\uparrow}(\vec{q},\w) = \chi^\uparrow_\mathrm{KS}(\vec{q},\w) + 
 \chi_\mathrm{KS}^\uparrow(\vec{q},\w)
\fHxc^{\uparrow\uparrow}(\vec{q},\w)
\chi^{\uparrow\uparrow}(\vec{q},\w).
\label{eq.Hspin}
\end{equation}
The above quantities are related to equation~\ref{eq.Ec} through
the simple substitutions
$\chi \rightarrow \chi^{\uparrow\uparrow}$ and 
$\chi_\mathrm{KS} \rightarrow \chi^{\uparrow}_\mathrm{KS}$.

Proceeding further requires the spin-polarized form of the 
Hartree-XC kernel $\fHxc^{\uparrow\uparrow}(\vec{q},\w)$.
To our knowledge, of the XC-kernels studied in this work
$\fHxc^{\uparrow\uparrow}$ has been derived only for the rALDA,
given as:\cite{Olsen2013}
\begin{equation}
\fHxc^{\uparrow\uparrow}(k) = \frac{4\pi}{k^2} - \left[ 2 \times \theta(k_c-k) \frac{4\pi}{k_c^2} + \theta(k-k_c) \frac{4\pi}{k^2} \right]
\label{eq.rALDA_spin}
\end{equation}
Equation~\ref{eq.rALDA_spin} differs from the spin-unpolarized rALDA
expression (equation~\ref{eq.ralda}) by a factor of two
in front of the part of the kernel corresponding to the ALDA,
reflecting the fact that the exchange interaction acts only
between electrons with the same spin.
Using equations~\ref{eq.Hspin}~and~\ref{eq.rALDA_spin} to 
calculate the correlation energy of the H atom yields a value
of 0.06~eV, compared to the exact value of zero and an RPA
value of -0.57~eV.

Taking the H and H$_2$ calculations together yields an rALDA atomization energy 
of 4.82~eV, which is within 0.1~eV of
the experimental value of 4.75~eV.\cite{Karton2006}
We note that the RPA benefits from substantial error 
cancellation and yields an atomization energy very close to experiment (4.74~eV, Table~\ref{tab.H2}).
However the H$_2$ molecule is a rather special case, and
the RPA usually demonstrates percentage errors of order 10\% in atomization
energies.\cite{Olsen2014}
The rALDA kernel corrects the correlation
energies of the individual H$_2$ and H systems and maintains
close agreement with the experimental atomization energy.

In Table~\ref{tab.H2} we also present the rALDA correlation energies using
the two-point density average, equation~\ref{eq.denssym}.
As found in bulk systems, the correlation energies calculated with the 
two-point density average are more negative ($\sim 0.06$~eV/electron) 
than those calculated with the symmetrized wavevector.
However
the agreement in atomization energies is better than 0.03~eV.
We find it encouraging that the symmetrized-wavevector approach
gives such similar results to the more intuitive two-point
density average when calculating the atomization energy.

We note that if we do not use the spin-polarized form of the kernel 
(equation~\ref{eq.rALDA_spin}), we find a correlation energy of -0.17~eV
for the H atom and an atomization energy of 4.37~eV.
This value is in significantly worse agreement with experiment than the 
RPA or even the LDA (4.89~eV), emphasizing the importance of
a rigorous treatment of spin.
An important direction for further study is the introduction
of spin-dependence into kernels derived from the spin-unpolarized HEG.

\section{Conclusions}
\label{sec.conclusions}

We have calculated the correlation energy of a test set
of 10 materials within the adiabatic-connection fluctuation-dissipation
formulation of density-functional theory (ACFD-DFT).
We used a hierarchy
of approximations for the exchange-correlation (XC) kernel $\fxc$, including
the random phase approximation (RPA, $\fxc = 0$), the recently-introduced renormalized kernels (rALDA),\cite{Olsen2013}
a kernel which satisfies the exact static limits of the electron gas (CDOP),\cite{Corradini1998} 
a model dynamical kernel (CPd)\cite{Constantin2007} and a kernel which diverges $\propto 1/k^2$
in the small-$k$ limit (JGMs).\cite{Trevisanutto2013}
In order to apply homogeneous kernels to inhomogeneous systems we 
applied a reciprocal space averaging scheme employing
wavevector symmetrization.\cite{Trevisanutto2013}
For each kernel and material pair we calculated the lattice constant
and bulk modulus, and compared our results to previous calculations
and experiment.\cite{Harl2010}

For all materials, including a nonzero $\fxc$ reduces the magnitude
of the correlation energy compared to the RPA by 0.1--0.5~eV per
electron.
This result mirrors the homogeneous electron gas (HEG), where the RPA correlation
energy is too negative by at least 0.3~eV over a wide range of
densities.\cite{Lein2000}
However the variation in correlation energy between each $\fxc$
is much smaller, on the scale of 0.01--0.1~eV per electron.
Encouragingly the correlation energies calculated with 
XC-kernels are found to lie very close 
to diffusion Monte Carlo data available for Si.\cite{Hood1998}
Furthermore, calculations with XC-kernels 
display faster basis-set convergence
than the RPA due to the suppression of high energy
plane-wave components of the Coulomb potential.

Considering lattice constants and bulk moduli, we found
only small variations between the RPA and different XC-kernels.
In particular, static XC-kernels that only satisfy the $k\rightarrow0$,
$\w=0$ limit of the HEG (rALDA, CP, CDOPs) produce very
similar results.
The structural properties calculated with the dynamical CPd kernel
are in better agreement with experiment in some cases (e.g.\ the 
bulk moduli of non-metallic systems), but the improvement is
not systematic (e.g.\ Na).
Satisfying the $k\rightarrow\infty$, $\w=0$ limit of the HEG (which adds
a local term to $\fxc$, e.g.\ the CDOP kernel) also yields
good agreement with experimental lattice constants, despite the kernel having a diverging
pair-distribution function.\cite{Lu2014}
Finally, the JGMs kernel predicts a reduction in lattice constants
and an increase in bulk moduli for non-metallic systems, bringing
the former into closer agreement with experiment.
The current study however cannot distinguish whether this behavior
is due to the general long-range $-\a/(4\pi R)$ character of the kernel,\cite{Botti2004}
or to the density-dependence of $\a$ specific to the JGMs model.\cite{Trevisanutto2013}

The ACFD-DFT scheme described here clearly involves a number of choices,
including (a) the approximation used to generate the noninteracting
response function $\chi_\mathrm{KS}$, (b) the form of $\fxc$ (including
spin-dependence) and (c) the
averaging scheme used to generalize a HEG XC-kernel to an inhomogeneous
system.
Fixing all factors except (b), as we have done here,
points us towards the essential properties of a model $\fxc$.
Based on the similar performance of the different XC-kernels, the
current work supports the idea that $\fxc$ should be kept as simple
as possible, i.e.\ be static, tend to a density-dependent constant at small $k$ 
and decay $\propto 1/k^2$
at large $k$.
In this respect the exchange-only rALDA kernel is attractive, since it 
scales simply with the coupling constant $\l$ and has good convergence
properties.
The introduction of additional computational expense and 
uncertainty associated with a dynamical kernel, 
a divergence $\propto 1 / k^2$ at $k=0$ or even a local term in $\fxc$
is difficult to justify based on the performance of the CPd, JGMs and CDOP kernels
for lattice constants and bulk moduli, although each kernel
was found to offer improved agreement with experiment in certain cases.

On the other hand, by focusing on the structural properties of 
bulk solids we have chosen systems
where the RPA already performs very well.
It is encouraging that the model XC-kernels can maintain this good
performance whilst correcting the magnitude of the correlation energy by several
eV per atom,
but arguably their real test lies in cases where the RPA is less successful.
Already the rALDA kernel has been found to improve the description
of atomization and cohesive energies\cite{Olsen2013,Olsen2014}
but a number of challenges remain, particularly in the description
of molecular dissociation.\cite{Harl2010,MoriSanchez2012,Henderson2010,Toulouse2009}
The framework described in the current study
provides the base for the application of a full range of kernels
to these more challenging systems.

\begin{acknowledgments}
We thank T.\ Olsen and J.J.\ Mortensen for useful advice and
discussions, and D.\ Lu for helpful correspondence.
The authors acknowledge support from the Danish Council for 
Independent Research's Sapere Aude Program, Grant No. 11-1051390.
\end{acknowledgments}


%
\end{document}